\documentclass[british,aps,twocolumn,pra,showpacs]{revtex4}
\usepackage[T1]{fontenc}
\usepackage[latin9]{inputenc}
\usepackage{babel}
\usepackage{amsmath}
\usepackage{amssymb}
\usepackage{graphicx}
\usepackage[unicode=true,pdfusetitle,
 bookmarks=true,bookmarksnumbered=false,bookmarksopen=false,
 breaklinks=false,pdfborder={0 0 1},backref=false,colorlinks=false]
 {hyperref}
\usepackage{breakurl}

\makeatletter
\@ifundefined{textcolor}{}
{%
 \definecolor{BLACK}{gray}{0}
 \definecolor{WHITE}{gray}{1}
 \definecolor{RED}{rgb}{1,0,0}
 \definecolor{GREEN}{rgb}{0,1,0}
 \definecolor{BLUE}{rgb}{0,0,1}
 \definecolor{CYAN}{cmyk}{1,0,0,0}
 \definecolor{MAGENTA}{cmyk}{0,1,0,0}
 \definecolor{YELLOW}{cmyk}{0,0,1,0}
}

\usepackage{bbm}

\usepackage{amsmath}
\DeclareMathOperator{\Tr}{Tr}

\newcommand\eqdef{\mathrel{\overset{\makebox[0pt]{\mbox{\normalfont\tiny def}}}{=}}}

\newcommand\eqquest{\mathrel{\overset{\makebox[0pt]{\mbox{\normalfont\tiny ?}}}{=}}}

\makeatother

\begin{document}

\title{Treating Time Travel Quantum Mechanically}

\author{John-Mark A. Allen}

\email{john-mark.allen@cs.ox.ac.uk}

\affiliation{Department of Computer Science, University of Oxford, Wolfson Building,
Parks Road, Oxford, OX1 3QD, UK}

\begin{abstract}
The fact that closed timelike curves (CTCs) are permitted by general
relativity raises the question as to how quantum systems behave when
time travel to the past occurs. Research into answering this question
by utilising the quantum circuit formalism has given rise to two theories:
Deutschian-CTCs (D-CTCs) and ``postselected'' CTCs (P-CTCs). In
this paper the quantum circuit approach is thoroughly reviewed, and
the strengths and shortcomings of D-CTCs and P-CTCs are presented
in view of their non-linearity and time travel paradoxes. In particular,
the ``equivalent circuit model''---which aims to make equivalent
predictions to D-CTCs, while avoiding some of the difficulties of
the original theory---is shown to contain errors. The discussion of
D-CTCs and P-CTCs is used to motivate an analysis of the features
one might require of a theory of quantum time travel, following which
two overlapping classes of new theories are identified. One such theory,
the theory of ``transition probability'' CTCs (T-CTCs), is fully
developed. The theory of T-CTCs is shown not to have certain undesirable
features---such as time travel paradoxes, the ability to distinguish
non-orthogonal states with certainty, and the ability to clone or
delete arbitrary pure states---that are present with D-CTCs and P-CTCs.
The problems with non-linear extensions to quantum mechanics are discussed
in relation to the interpretation of these theories, and the physical
motivations of all three theories are discussed and compared.
\end{abstract}
\maketitle

\section{Introduction\label{sec:Introduction}}

There are few areas of physics in which one confronts the idea of
time travel to the past. When it is discussed, time travel is often
associated with particular ways of thinking about quantum mechanics
and quantum field theory. One often hears the view that in quantum
teleportation, for example, the teleported information travels back
in time to the point at which entanglement was created, before proceeding
forward in time to the recipient \cite{Penrose98,Jozsa98,Jozsa04,Timpson06}.
Another example might be remarks about antiparticles in quantum field
theory being akin to particles ``travelling backwards in time''.
The time travel considered in this work is rather different. Here
it is assumed that time travel into the past is a possible physical
process. The question is then: How must quantum mechanics be modified
to account for time travel? The motivation for this comes not from
within quantum theory itself, but rather from general relativity and
the existence of spacetime solutions containing closed timelike curves
(CTCs).

CTCs are paths through exotic spacetimes along which massive particles
may travel, apparently forwards through time, only to return to their
own past. As such, they represent a mechanism for time travel to the
past. The discovery that the Einstein field equations of general relativity
permit spacetime solutions that contain CTCs \cite{Godel49,Bonnor80,Gott91}
came as something of a surprise \cite{Einstein98}. Many believe that
such spacetimes containing CTCs should not occur in nature \cite{Hawking92,Morris88};
however, there is no consensus on the possibility of precluding such
solutions. Spacetimes containing CTCs are globally non-hyperbolic
and therefore are incompatible with standard relativistic quantum
field theory \cite{Birrell82}. Such spacetimes typically contain
no Cauchy surfaces \cite{Wald10} and so the initial value problem
is not well defined, making the analysis of matter in such spacetimes
tricky.

Deutsch \cite{Deutsch91} began a programme for investigating quantum
mechanics along CTCs that uses the quantum circuit formalism of quantum
computation \cite{Nielsen00}. In this approach a circuit is used
to ferry qubit states along classical paths between localised regions
of spacetime where they evolve and interact according to unitary quantum
theory; the relativistic effects of CTCs are present only in the ability
to send qubits back in time. In this way, the resulting theories abstract
away from the specific mechanism of time travel, and may, therefore,
be applicable to any such mechanism, be it TARDIS, DeLorean, or CTC.

Deutsch's theory of quantum mechanics with time travel has come to
be known as the theory of D-CTCs. A second theory of quantum mechanics
with time travel based on the same quantum circuit approach has come
to be known as the theory of P-CTCs, since it uses the mathematics
of postselection \cite{Svetlichny09,Lloyd11a,Lloyd11b}. P-CTCs are
motivated by ``teleporting'' quantum states into the past and are
equivalent to being able to postselect on quantum measurement outcomes.

Science fiction is rife with naïve classical examples of time travel
to the past, almost inevitably leading to various paradoxes. To some
extent, both D-CTCs and P-CTCs mitigate such paradoxes. However, both
theories still contain undesirable features that may also be pathological
or paradoxical.

The evolution of a state in a region involving time travel is expected
to be both non-unitary \cite{Hawking95,Anderson95,Fewster95} and
non-linear \cite{Hartle94}. Non-linearity, if present, fundamentally
changes the structure of quantum theory. The standard proofs of most
central theorems in quantum mechanics (including no-signalling, no-cloning,
generalised uncertainty principles, and indistinguishably of non-orthogonal
states) depend on the linearity of the theory. It is therefore of
little surprise that: both D-CTCs and P-CTCs can distinguish non-orthogonal
states with a single measurement and (in certain ontologies) signal
faster than light; D-CTCs can clone arbitrary pure states; and P-CTCs
can delete arbitrary states. Linearity is also necessary for the equivalence
of proper and improper mixed states in ordinary quantum theory \cite{dEspagnat06,Brun12}
and the ability of quantum theory to be compatible with many different
interpretations and ontologies \cite{Saunders10}. Non-linear extensions
of quantum theory, such as D-CTCs and P-CTCs, therefore, require careful
examination of any ontological assumptions as the predictions of the
theories generally depend on them.

Such issues surrounding theories of time travel in quantum circuits
are subtle, and different theories give different perspectives. As
well as containing undesirable features, both D-CTCs and P-CTCs are
more difficult to interpret than standard quantum mechanics. Since
they are constructed from the abstract quantum circuit formalism,
any interpretational assumptions behind them are unclear. One may
therefore ask if there could be a theory with fewer undesirable features.
Could such a theory have a more robust physical motivation?

In this paper, the first of these questions is answered squarely in
the affirmative. Several alternative theories of time travel in quantum
mechanics, falling into two overlapping classes, are presented. One
such theory, called the theory of T-CTCs (since it uses transition
probabilities), is thoroughly developed and seen to have far fewer
problematic features than either D-CTCs or P-CTCs; in particular,
cloning and deleting are impossible and non-orthogonal states cannot
be distinguished with a single measurement. The second question is
more subjective in its nature, but the theories presented here all
have physical motivations that may be, depending one's prejudice,
at least as compelling as those of D-CTCs and P-CTCs.

This strongly suggests that the question as to how quantum mechanics
behaves in the presence of time travel, and specifically CTCs, is
far from decided. With so many potential answers, different insights
and criteria are needed to convincingly select one theory to the exclusion
of the others.

Sections \ref{sec:Modelling-Time-Travel} and \ref{sec:D-CTCs-and-P-CTCs}
of this paper largely consist of a review of time travel in the quantum
circuit formalism, but they do also contain some new results and define
the notation and concepts used in the later sections. In Sec. \ref{sec:Modelling-Time-Travel}
the framework of quantum circuits with time travel is developed and
classical time travel discussed. In Sec. \ref{sec:D-CTCs-and-P-CTCs}
the theories of D-CTCs and P-CTCs are presented with a discussion
of their features. As part of the development of D-CTCs, the equivalent
circuit model---a proposed alternative model of D-CTCs \cite{Ralph10,Ralph12}---is
shown to contain errors. Readers familiar with these theories may
want to skim through this section rather rapidly. The meat of the
paper is in Sec. \ref{sec:New-Theories} where certain desirable features
of any theory of quantum time travel are identified, and used to find
two classes of new theories; from these new theories that of T-CTCs
is selected and fully developed. Finally, in Sec. \ref{sec:Discussion},
the theory of T-CTCs is compared to the previous theories, focussing
on their motivation, shortcomings, and successes.

\section{Modelling Time Travel with Quantum Circuits\label{sec:Modelling-Time-Travel}}

\subsection{The Standard Form of Quantum Circuits with Time Travel\label{sub:Standard-Form-of}}

The quantum circuit approach to time travel is based on a particular
form of qubit circuit introduced by Deutsch \cite{Deutsch91}. This
convenient building block, from which all other circuits involving
time travel can be built, is here called the \emph{standard form circuit}.
It allows different circuits and theories to be concisely specified.

The advantage of the quantum circuit model in this context is that
quantum evolution of qubit states is separated from spatial motion.
Quantum interactions described by unitary gates are assumed to only
occur in small, freely falling, non-rotating regions of spacetime
so that they obey non-relativistic quantum mechanics. To include time
travel to the past in such a quantum circuit model is therefore equivalent
to saying that the classical paths qubits take between gates are allowed
to go back in time. The restriction to circuits of qubits may seem
unphysical but it is convenient and expected to be computationally
universal \cite{Deutsch91}; circuits with systems other than qubits
may also be constructed by analogy. Within this approach, different
theories are then defined by their behaviour when sending qubits back
in time. In this paper only finite-dimensional Hilbert spaces are
considered by considering only finite numbers of qubits.

\begin{figure}
\begin{centering}
\includegraphics[width=8cm]{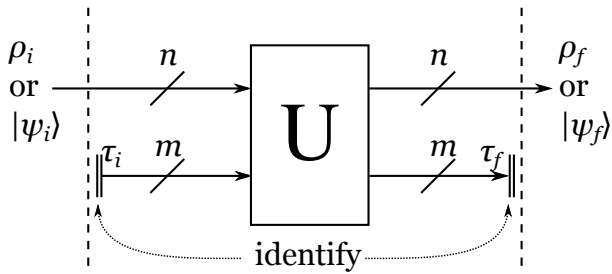}
\par\end{centering}

\protect\caption{Schematic diagram for the standard form circuit described in the text.
The $n$ CR and $m$ CV qubits are shown entering the gate labelled
by unitary $U$. The double bars represent the time travel event and
may be thought of as two depictions of the same spacelike hypersurface,
thus forming a CTC. The dashed lines represent the spacelike boundaries
of the region in which time travel takes place; CV qubits are restricted
to that region.}

\label{fig:standardform}
\end{figure}

The standard form circuit is a circuit of $n+m<\infty$ qubits, a
single time travel event that occurs in a localised spacetime region
\footnote{The corresponding statement in the quantum circuit model is that there
exist qubit paths in the circuit that do not go back in time. This
is not strictly necessary, but it simplifies the discussion. Even
if the universe contains CTCs that are not localised in this way,
one might reasonably expect any theory to be extendible to a universe
where they all are.%
}, and a single unitary quantum interaction $U$. This paper does not
consider situations in which the time travel region is not localised
in spacetime. The $m$ qubits that travel back in time are called
the \emph{chronology violating }(CV)\emph{ qubits}, which arrive from
their own future in the state $\tau_{i}$ and after the interaction
are said to be in the state $\tau_{f}$. The remaining $n$ qubits
are called the \emph{chronology respecting }(CR)\emph{ qubits, }which
arrive from the unambiguous past in the state $\rho_{i}$ and emerge
into the unambiguous future in the state $\rho_{f}$. Therefore, a
standard form circuit is completely specified by values for $n$,
$m$, and $U$, while a theory for quantum mechanics with time travel
is a specification of $\rho_{f}$ given $U$ and $\rho_{i}$. A generic
standard form circuit is illustrated in Fig. \ref{fig:standardform}.

In cases where the CV system is initially known to be in a pure state
it may be written as $|\phi\rangle$ so that $\tau_{i}=|\phi\rangle\langle\phi|$.
Similarly, in cases where the CR system input or output states are
known to be pure, they may be written $|\psi_{i,f}\rangle$ so that
$\rho_{i,f}=|\psi_{i,f}\rangle\langle\psi_{i,f}|$. In Sec. \ref{sub:Types-of-Mixed}
it is seen that in all models of quantum time travel considered here
one may always purify such that $\rho_{i}$ is pure.

All theories of quantum time travel presented in this paper make one
additional assumption: Before the interaction $U$ the CR and CV systems
are not entangled so that the state of all $n+m$ qubits is described
by the product state $\rho_{i}\otimes\tau_{i}$ \cite{DeJonghe10}.
In Ref. \cite{Politzer94} it is argued that this is unreasonable
as the causal past of the CV qubits contains $\rho_{i}$ and entanglement
should therefore be possible, but that possibility is not considered
here because alternative assignment methods all have undesirable features.
The problem is one of finding an assignment procedure that gives a
joint state $\omega$ to $\mathcal{H}_{\textsc{CR}}\otimes\mathcal{H}_{\textsc{CV}}$
given only a state $\rho_{i}$ on $\mathcal{H}_{\textsc{CR}}$. The
procedure used here---where $\omega=\rho_{i}\otimes\tau_{i}$---is
the only one for which: $\omega$ is always positive, $\Tr_{\textsc{CV}}\omega=\rho_{i}$,
and mixtures are preserved \cite{Alicki95}.

\subsection{Time Travel Paradoxes and the Classical Model\label{sub:Time-Travel-Paradoxes}}

In order define the possible types of paradox in time travel it is
useful to leave quantum mechanics to one side briefly and concentrate
on classical time travel. Consider a classical version of the standard
form circuit, with classical states $\tilde{\rho}_{i,f}$ and $\tilde{\tau}_{i,f}$
and some classical dynamical evolution $\tilde{U}$ replacing their
quantum counterparts. An \emph{ontic state} of a system is a complete
physical state according to some theory, while an \emph{epistemic
state} is a probability distribution over ontic states reflecting
some observer's uncertainty of the ontic state.

The standard way of introducing time travel into classical theories
is to impose a consistency condition on states that go back in time;
that is, $\tilde{\tau}_{i}=\tilde{\tau}_{f}\eqdef\tilde{\tau}$, where
$\tilde{\tau}$ is an ontic state. In other words, the ontic state
that emerges in the past is required to be the same one that left
from the future. With knowledge of an ontic state $\tilde{\rho}_{i}$
and $\tilde{U}$, a consistent $\tilde{\tau}$ can be deduced, from
which $\tilde{\rho}_{f}$ may be calculated.

There are two distinct types of paradox that may arise when considering
time travel to the past classically: paradoxes of \emph{dynamical
consistency }and paradoxes of \emph{information}. In this paper a
slightly unconventional view on paradoxes is taken whereby a computation,
which algorithmically produces unambiguous output from an input, is
never paradoxical. So a theory that predicts ``absurdly'' powerful
communication or computational abilities will not be called paradoxical
because of them, even if those abilities make the theory appear unreasonable
or hard to accept. The ``absurd'' ability of time travel has been
assumed, so one should expect some ``absurd'' conclusions.

A dynamical consistency paradox is a situation in which a consistent
history of events is not possible \cite{Deutsch91,Lloyd11a,Lloyd11b}.
The ``grandfather paradox'', in which a grandpatricidal time traveller
goes back to kill their infant grandfather, is the usual example of
such a paradox. Dynamical consistency paradoxes arise when a theory
fails to specify any valid final state from some initial state and
evolution. Classically, this occurs because the consistency condition
is rendered unsatisfiable either for a particular set of possible
$\tilde{\rho}_{i}$ or for all $\tilde{\rho}_{i}$ %
\footnote{It is worth noting that, although these paradoxes appear to be possible
in classical models of physics, such situations are quite difficult,
if not impossible, to construct in classical models with continuous
state spaces. Reference \cite{SEP-Time-Travel-Phys} is a useful introduction
to such issues.%
}. There are three options to avoid these paradoxes when they arise
in a theory: disallow time travel in that theory; only allow time
travel with certain interactions that avoid these paradoxes; or enforce
retrospective constraints on the initial conditions. The first option
is contrary to the object of study and is not considered. Retrospective
constraints are insufficient to prevent paradox when interactions
exist that preclude any consistent input; therefore, such interactions
must be disallowed from any theory of time travel. Reference \cite{Deutsch91}
and references therein discuss whether such retrospective constraints
are acceptable. 

\begin{figure}
\begin{centering}
\includegraphics[width=6cm]{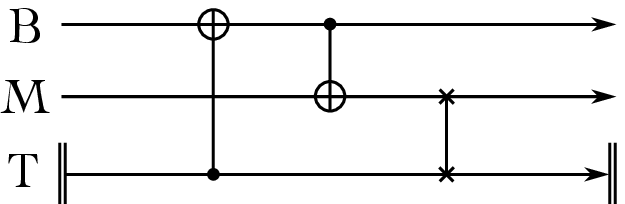}
\par\end{centering}

\protect\caption{Circuit diagram for the unproven theorem circuit of Ref. \cite{Lloyd11a}
in the standard form. The book qubit is labelled $B$, mathematician
labelled $M$, and time-traveller labelled $T$. The input to the
circuit is taken to be the pure state $|0\rangle_{B}|0\rangle_{M}$.
The notation used for the gates is the standard notation as defined
in Ref. \cite{Nielsen00} and describes a unitary $U=\textsc{SWAP}_{MT}\textsc{CNOT}_{BM}\textsc{CNOT}_{TB}$.}
\label{fig:unproventheorem}
\end{figure}

Information paradoxes are situations with consistent dynamics in which
information appears that has no source; \emph{viz.}, the information
has not been computed. The prototypical example is the \emph{unproven
theorem} paradox: A mathematician reads the proof of a theorem, travels
back in time and writes the proof down in a book; the younger version
of the same mathematician learns this proof from said book and subsequently
goes back in time in order to fulfil their destiny of writing the
proof down \cite{Deutsch91,Lloyd11a,Lloyd11b,Lloyd11c,Ralph11a}.
The proof appears to have emerged from nowhere. A quantum circuit
that models this paradox was given in Ref. \cite{Lloyd11a} and is
illustrated in Fig. \ref{fig:unproventheorem}. The circuit consists
of a CR ``book'' qubit $B$, a CR ``mathematician'' qubit $M$,
and a CV ``time traveller'' qubit $T$. In the standard form this
circuit has $n=2$, $m=1$, and $U=\textsc{SWAP}_{MT}\textsc{CNOT}_{BM}\textsc{CNOT}_{TB}$.
The input is taken to be the pure state $|0\rangle_{B}|0\rangle_{M}$.
The interpretation is that the time traveller writes the theorem in
the book ($\textsc{CNOT}_{TB}$), the mathematician then reads the
book ($\textsc{CNOT}_{BM}$), and finally the mathematician and time
traveller swap places ($\textsc{SWAP}_{MT}$) so that the mathematician
may go back in time to write the theorem. Using $N$ copies of the
circuit ($n=2N$ and $m=N$) enables the mathematician to encode a
theorem in an $N$-bit binary string.

Under the definitions above, information paradoxes arise if and only
if a theory contains a \emph{uniqueness ambiguity}: The theory specifies
more than one final state given some initial state and evolution,
but fails to give probabilities for each possibility. The reason for
this is simple: Any other dynamically consistent evolution is counted
as a computation and information paradoxes are defined to be exactly
those dynamically-consistent situations where an uncomputed output
is produced. For example, in the classical unproven theorem paradox
there are many time travelling states $\tilde{\tau}$ compatible with
the consistency condition, each producing a different $\tilde{\rho}_{f}$---one
$\tilde{\tau}$ produces a theorem answering $\mathsf{P}\eqquest\mathsf{NP}$,
another answering $\mathsf{BPP}\eqquest\mathsf{BQP}$, many others
where the ``theorem'' is nonsense, etc. It should be noted that
this conclusion on the equivalence of time travel situations with
uniqueness ambiguity and those with information paradoxes is not one
that is generally accepted, but it follows from the definitions used
here as described above. Readers preferring different definitions
may replace each further instance of ``information paradox'' with
``uniqueness ambiguity'' without affecting the meaning.

Some authors prefer a wider definition of paradox, in which time travel
circuits constructed such that the only consistent evolution reveals
a fixed point of some given function is counted as an information
paradox \cite{Deutsch91,Lloyd11c}. Such a circuit uniquely produces
the solution to a problem that is hard if $\mathsf{P}\neq\mathsf{NP}$
\cite{Garey79} very rapidly, and as such is counted as a very powerful
computation rather than a paradox as defined here. This is not to
claim that such processes are necessarily reasonable, but rather to
reflect that they are qualitatively different from the non-unique
situations equivalent to information paradoxes by the definitions
of this paper. Nonetheless, it is worth noting that both D-CTCs and
P-CTCs are also capable of solving for fixed points in this way.

By contrast, uniqueness ambiguity is required to cause information
paradoxes such as the unproven theorem paradox. If there were no uniqueness
ambiguity then any information that appears as a result of the time
travel circuit is uniquely specified by the structure of, and input
to, the circuit \cite{Lloyd11c}. It is therefore reasonable to say
that this circuit is algorithmically computing as instructed.

\section{D-CTCs and P-CTCs\label{sec:D-CTCs-and-P-CTCs}}

\subsection{D-CTCs\label{sub:D-CTCs}}

One may construct the theory of D-CTCs rapidly by following Ref. \cite{Wallman12}
and assuming that reduced density operators are ontic states %
\footnote{Note that this is not the line of reasoning that Deutsch originally
took to develop the theory, but it is a rapid and instructive route.%
}. Just as in the classical model, a consistency condition $\tau_{i}=\tau_{f}\eqdef\tau$
should be imposed on the ontic time travelling states. In the standard
form this implies 
\begin{equation}
\tau=\mathsf{D}(\tau)\eqdef\Tr_{\textsc{CR}}[U(\rho_{i}\otimes\tau)U^{\dagger}].\label{eq:d-ctccc}
\end{equation}
Since the partial trace is used to separate CR and CV systems a consistent
approach then suggests that the final state should be 
\begin{equation}
\rho_{f}=\Tr_{\textsc{CV}}[U(\rho_{i}\otimes\tau)U^{\dagger}].\label{eq:d-ctcem}
\end{equation}
Essentially, the ingredient that makes D-CTCs different to ordinary
unitary quantum mechanics is the map

\begin{equation}
U\left(\rho_{i}\otimes\tau\right)U^{\dagger}\rightarrow\rho_{f}\otimes\tau,\label{eq:d-ctcdynamicalchange}
\end{equation}
which replaces the general quantum state after $U$ with the product
of its reduced density operators. The point at which the non-quantum
map described in Eq. (\ref{eq:d-ctcdynamicalchange}) is supposed
to occur is not defined, and so there is a \emph{dynamical ambiguity}
in the theory. However, this ambiguity is entirely without observable
consequence since all local observations on separated parts of a bipartite
system are entirely dictated by the reduced density operators, and
Eq. (\ref{eq:d-ctcdynamicalchange}) simply maps a bipartite system
onto the product of its reduced density operators \cite{Bacon04}.

Equations (\ref{eq:d-ctccc}) and (\ref{eq:d-ctcem}) define the theory
of D-CTCs in its barest form: To find $\rho_{f}$ given $\rho_{i}$
and $U$, one solves Eq. (\ref{eq:d-ctccc}) to obtain $\tau$ and
then evaluates Eq. (\ref{eq:d-ctcem}). Note that the equation of
motion (\ref{eq:d-ctcem}) is both non-linear and non-unitary for
general $\rho_{i}$ and $U$.

The superoperator $\mathsf{D}$ is the Stinespring dilation form \cite{Stinespring55,Nielsen00}
of an ordinary linear quantum channel. Schauder's fixed point theorem
\cite{Schauder30,Tychonoff35,Zeidler85} guarantees that any trace-preserving
quantum channel between density operators has at least one fixed point
$\tau$. Therefore, any D-CTC with any unitary and input state can
be solved for an output state $\rho_{f}$ and D-CTCs are not vulnerable
to dynamical consistency paradoxes that are present classically, despite
being based on a very similar consistency condition. Direct and instructive
proofs of this can be found in Refs. \cite{Deutsch91,Blume-Kohout08}.

\subsubsection{The Uniqueness Ambiguity in D-CTCs\label{sub:Non-uniqueness-Ambiguity-in}}

Whilst a solution of Eq. (\ref{eq:d-ctccc}) for $\tau$ always exists,
this solution is not always unique; therefore, D-CTCs contain the
same uniqueness ambiguity present in classical time travel. As such,
they are vulnerable to unproven theorem paradoxes. Most theorems involving
D-CTCs in the literature use circuits that permit only a single solution
for $\tau$ and thereby avoid this ambiguity, but it is present in
general.

Consider the unproven theorem circuit of Sec. \ref{sub:Time-Travel-Paradoxes}.
Solving Eq. (\ref{eq:d-ctccc}), one finds a family of CV states,
$\tau=\alpha|0\rangle\langle0|+(1-\alpha)|1\rangle\langle1|$, and
a corresponding family of output states,
\begin{equation}
\rho_{f}=\alpha|0\rangle_{B}\langle0|\otimes|0\rangle_{M}\langle0|+(1-\alpha)|1\rangle_{B}\langle1|\otimes|1\rangle_{M}\langle1|,
\end{equation}
for $0\leq\alpha\leq1$. As such, there is a continuous one-parameter
family of possible solutions in the D-CTC case and it is seen that
D-CTCs are vulnerable to the unproven theorem paradox.

In Ref. \cite{Deutsch91} the \emph{maximum entropy rule} was suggested
to resolve this ambiguity. The rule states that one should choose
the consistent $\tau$ with the greatest von Neumann entropy $S(\tau)$.
This always specifies a unique $\tau$ due to linearity of $\mathsf{D}$
and concavity of $S$ \cite{DeJonghe10} and was motivated by a desire
to limit the ability of D-CTCs to solve fixed point problems in the
manner discussed in Sec. \ref{sub:Time-Travel-Paradoxes}. However,
since the maximum entropy rule has not been universally accepted and
other possible principles exist \cite{Politzer94,DeJonghe10}, in
this paper it is viewed as a non-essential extension to the theory
of D-CTCs. In the case of the unproven theorem circuit discussed above,
the maximum entropy rule would uniquely specify $\alpha=\frac{1}{2}$.
The interpretation of the unproven theorem circuit for this solution
is that the ``book'' $B$ contains an equal mixture of all possible
``theorems'' (most of which will be nonsense) and is useless, so
this also resolves the unproven theorem paradox for this example. 

By allowing for some noise along the path of the CV system, it may
be possible to avoid the uniqueness ambiguity. Note that, as a consistency
condition, Eq. (\ref{eq:d-ctccc}) must be exactly, rather than approximately,
satisfied. In order to incorporate noise into a D-CTC circuit, one
must modify the circuit, including a new quantum channel $\mathsf{N}$
responsible for the noise. Equation (\ref{eq:d-ctccc}), is therefore
modified to $\tau=\mathsf{N}(\mathsf{D}(\tau))$, where $\mathsf{D}$
remains the quantum channel associated with the noiseless circuit.
Suppose that the noise $\mathsf{N}$ is modelled with a depolarising
channel \cite{Nielsen00} such that depolarisation occurs with probability
$0<p<1$. The consistency condition then becomes
\begin{equation}
\tau=\frac{p}{2^{m}}\mathbbm{1}+(1-p)\mathsf{D}(\tau).\label{eq:d-ctcdecoherence}
\end{equation}
Now suppose that two density operators satisfying Eq. (\ref{eq:d-ctcdecoherence})
differ by $\Delta\tau$. By linearity of $\mathsf{D}$, $\Delta\tau=(1-p)\mathsf{D}(\Delta\tau)$
and therefore the solution to Eq. (\ref{eq:d-ctcdecoherence}) can
only be non-unique if $(1-p)^{-1}$ is an eigenvalue of $\mathsf{D}$.
In a finite-dimensional Hilbert space, such as those of finitely many
qubits considered here, $\mathsf{D}$ has a discrete spectrum of eigenvalues
\cite{Wolf10}, while $(1-p)^{-1}$ takes arbitrary values from a
continuum. Therefore, $(1-p)^{-1}$ will not generally be an eigenvalue
of $\mathsf{D}$ so the addition of an arbitrary depolarising channel
specifies $\tau$ uniquely. It therefore appears that the addition
of noise to the CV system may solve the uniqueness ambiguity without
need for the maximum entropy rule. The conjecture that by considering
general noise on the CV system the uniqueness ambiguity may be avoided
is supported by the observation in Ref. \cite{Politzer94} that the
subspace of unitaries that give rise to the uniqueness ambiguity with
D-CTCs is of measure zero. However, the uniqueness ambiguity is still
present in the theory, even if it disappears under small noise.

\subsubsection{Interpretation of D-CTCs\label{sub:Interpretation-of-D-CTCs}}

D-CTCs were introduced above by assuming that density operators are
ontic, and reasoning that one should require consistency of ontic
states in time travel. This interpretation is favoured in Ref. \cite{Wallman12},
in which D-CTCs are reviewed from an epistemic perspective and found
to be inconsistent.

While issues with such an epistemic interpretation were recognised
by Deutsch, he did not present any ontological assumptions from which
the theory is derived \cite{Deutsch91}. He did, however, favour an
interpretation in terms of Everettian quantum theory \cite{DeWitt73,Saunders10}.
This provides an interpretation for the disentanglement process in
D-CTCs expressed in Eq. (\ref{eq:d-ctcdynamicalchange}): When a system
traverses a D-CTC, it passes between Everettian branches into a different
``world'' from the one from which it left.

As presented, the entanglement breaking map Eq. (\ref{eq:d-ctcdynamicalchange})
has similarities with the ``collapse'' of the state on measurement
in ordinary quantum mechanics. Whilst it is entirely unobservable,
on an interpretational level this disentanglement process needs to
be accounted for in a way similar to the apparent post-measurement
state collapse of standard quantum mechanics. The Everettian view
of Ref. \cite{Deutsch91} is one way of approaching this.

\subsubsection{The Equivalent Circuit Model}

The equivalent circuit model of D-CTCs is a proposed alternative model
for D-CTCs \cite{Ralph10,Ralph12} that attempts to remove two ambiguities
from the theory: the uniqueness ambiguity and the treatment of mixed
states by the theory (Sec. \ref{sub:Mixed-states-and}). In this section
it is demonstrated that, in its standard presentation, the equivalent
circuit model fails to generally reproduce the theory of D-CTCs and
that the proof given for the resolution of the uniqueness ambiguity
contains errors.

\begin{figure}
\begin{centering}
\includegraphics[width=1\columnwidth]{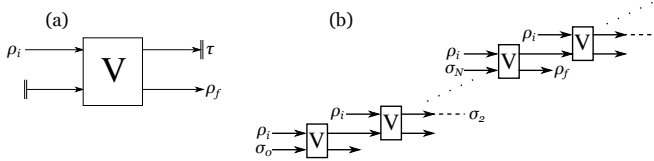}
\par\end{centering}

\protect\caption{The equivalent circuit model. (a) A circuit equivalent to the standard
form of Fig. \ref{fig:standardform} with $V=\textsc{SWAP}U$. (b)
The same circuit ``unwrapped'' in the equivalent circuit model.
There is an infinite ladder of unwrapped circuits, each with the same
input state $\rho_{i}$. To start the ladder, an initial CV state
$\sigma_{0}$ is guessed. The CV state of the $N$th rung of the ladder
is $\sigma_{N}$. The output state $\rho_{f}$ is taken after a number
$N\rightarrow\infty$ of rungs of the ladder have been iterated.}

\label{fig:equivalentcircuit}
\end{figure}

The equivalent circuit model is defined for standard form circuits
and is outlined in Fig. \ref{fig:equivalentcircuit}. The unitary
$V$ differs from the $U$ found in the corresponding standard form
circuit by only a unitary which switches the CV and CR Hilbert spaces.
Unitarity is restored by ``unwrapping'' the circuit so that each
time a system goes back in time, there is a new copy of the circuit,
as shown in Fig. \ref{fig:equivalentcircuit}(b). These copies of
the circuit form a infinite ``ladder'', the ``rungs'' of which
are CV qubit paths. The unwrapped circuit, therefore, has infinitely
many outputs from which one, infinitely far up the ladder, is selected.

Following Ref. \cite{Ralph10}, the standard way of calculating with
these unwrapped equivalent circuits is to find $\tau$ by starting
the infinite ladder of circuits with a guess for $\tau$, which is
here called $\sigma_{0}$ %
\footnote{In Ref. \cite{Ralph10} $\sigma_{0}=\rho_{i}$ is used.%
}. The state of the CV system on the $N$th ``rung'' of the ladder
of circuits is given by
\begin{equation}
\sigma_{N}=\mathsf{D}^{N}(\sigma_{0}),\quad\mathsf{D}(\sigma)=\Tr_{2}[V(\rho_{i}\otimes\sigma)V^{\dagger}]\label{eq:equivalentcircuit}
\end{equation}
where the partial trace is taken over the final $n$ qubits in the
tensor product {[}lower arm in each unitary in Fig. \ref{fig:equivalentcircuit}(b){]}.
Note that $\mathsf{D}$ is exactly as defined in Eq. (\ref{eq:d-ctccc}).
The claim of the equivalent circuit model is that, since $\mathsf{D}$
always has fixed points, iterating $\mathsf{D}$ as in Eq. (\ref{eq:equivalentcircuit})
will always cause $\sigma_{N}$ to converge to a fixed point of $\mathsf{D}$;
\emph{i.e.,} $\tau=\lim_{N\rightarrow\infty}\sigma_{N}$: ``Deutsch
showed that such a fixed point always exists, so the assumed convergence
is guaranteed'' \cite{Ralph10}. This is not generally correct and
counter examples can be constructed. For example, suppose that $n=m=1$
and $V=(X_{1}\otimes\mathbbm{1}_{2})(\textsc{CNOT}_{1,2})(\textsc{SWAP})$,
with the control qubit for $\textsc{CNOT}$ being the upper arm of
the unitary as illustrated in Fig. \ref{fig:equivalentcircuit} and
$X_{1}$ the Pauli $X$ unitary applied to the same qubit. Starting
with $\rho_{i}=\sigma_{0}=|0\rangle\langle0|$ results in subsequent
$\sigma_{N}$ alternating between $|1\rangle\langle1|$ and $|0\rangle\langle0|$
and the corresponding $\rho_{f}$ alternating between $|0\rangle\langle0|$
and $|1\rangle\langle1|$, never converging. So repeated action of
$\mathsf{D}$ does not generally reproduce the consistency condition
as claimed.

Another claim of the equivalent circuit model is that, given arbitrarily
small decoherence along the CV system, the maximum entropy rule is
reproduced and is equivalent to simply choosing $\sigma_{0}\propto\mathbbm{1}$.
Notwithstanding the comments of the previous section and the argument
above, it is now shown that the argument justifying this claim presented
in the appendix of Ref. \cite{Ralph10} is mistaken.

Suppose that repeated action of $\mathsf{D}$ does cause convergence
to a fixed point and write $\tau(\rho_{i},\sigma_{0})$ for the fixed
point converged to when the iteration starts from $\sigma_{0}$ and
the input state is $\rho_{i}$. In Eq. (A3) of Ref. \cite{Ralph10}
the entire process of convergence to a fixed point is written as a
Kraus decomposition with Kraus operators $\{E_{j}\}_{j}$ \cite{Kraus71,Nielsen00}.
The reasoning then follows 
\begin{eqnarray}
\tau(\rho_{i},\sigma_{0}) & = & \lim_{N\rightarrow\infty}\mathsf{D}^{N}(\sigma_{0})=\sum_{j}E_{j}\sigma_{0}E_{j}^{\dagger}\nonumber \\
\sum_{j}E_{j}\sigma_{0}E_{j}^{\dagger} & = & \sum_{j,k}E_{j}E_{k}\sigma_{0}E_{k}^{\dagger}E_{j}^{\dagger}\\
\sum_{j,k}E_{j}E_{k}^{\dagger}E_{k}\sigma_{0}E_{j}^{\dagger} & = & \sum_{j,k}E_{j}E_{k}\sigma_{0}E_{k}^{\dagger}E_{j}^{\dagger}\nonumber 
\end{eqnarray}
where the second line follows since $\tau$ is a fixed point. It is
then claimed that this implies that $[E_{j}\sigma_{0},\, E_{j}^{\dagger}]=0$.
This is not true generally; for example, it would imply that $\tau(\rho_{i},\sigma_{0})=\sigma_{0}$
for all $\sigma_{0}$ and therefore, by the assumption that $\tau(\rho_{i},\sigma_{0})$
is a fixed point, that every $\sigma_{0}$ is a fixed point of $\mathsf{D}$.
Thus, the claimed proof is, at best, incomplete. 

It may be possible to salvage the equivalent circuit model in such
a way that performs as advertised, but the above discussion shows
that the model and its proofs need careful revision before it does
so.

\subsubsection{Computation with D-CTCs}

In Ref. \cite{Aaronson09} it was proved that D-CTCs are able to compute
any problem in the complexity class $\mathsf{PSPACE}$ in polynomial
time, meaning that D-CTCs are thought to be vastly more powerful than
even quantum computers, which can calculate problems in $\mathsf{BQP}$.
In the same paper, there is a proof that classical computers with
time travel are also able to compute any problem in $\mathsf{PSPACE}$
in polynomial time. However, it should be noted that the model of
classical time travel used in that proof is different from the model
of Sec. \ref{sub:Time-Travel-Paradoxes}. In particular, in Ref. \cite{Aaronson09}
classical time travel does not impose a consistency condition on ontic
classical states $\tilde{\tau}$, but rather on classical epistemic
states of the time travelling system; \emph{viz.}, probability distributions
over possible $\tilde{\tau}$. Such a model for classical time travel
does not seem physically motivated, as classical probability distributions
are assumed to be entirely due to observer ignorance rather than anything
intrinsically physical. It does, however, have the pleasing computational
property of not suffering dynamical consistency paradoxes. Therefore,
claims that classical computers with time travel have the power of
$\mathsf{PSPACE}$, or that they are equivalent to quantum computers
with time travel, should be taken with caution as they are based on
an unusual model of classical time travel.

It is worth noting that D-CTCs can also produce discontinuous maps
from $\rho_{i}$ to $\rho_{f}$ \cite{DeJonghe10}. For all practical
purposes, the theory loses predictive power when the input state is
close to one of these discontinuities in a similar way that the onset
of chaos causes loss of predictive power about a classical dynamical
system.

\subsection{P-CTCs}

The theory of P-CTCs is due to Svetlichny \cite{Svetlichny09}, inspired
by Coecke's work on diagrammatic approaches to quantum mechanics %
\footnote{See footnote 1 of \cite{Svetlichny09}. The diagrammatic approach
was also instrumental in helping to find some of the new theories
presented here.%
}, and Lloyd \emph{et al. }\cite{Lloyd11a,Lloyd11b}, based on the
unpublished work of Bennett and Schumacher \cite{Bennett05} and inspired
by the Horowitz-Maldacena final state condition of black hole evaporation
\cite{Horowitz04}. The core idea is to use the ordinary quantum mechanical
teleportation protocol as a basis for teleporting qubits into the
past. Reference \cite{Brun12} contains an accessible introduction.

\begin{figure}
\centering{}\includegraphics[width=8cm]{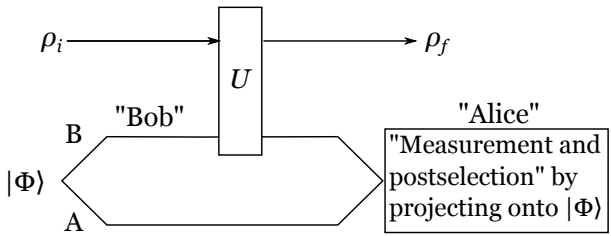}\protect\caption{Schematic diagram illustrating the protocol defining the action of
P-CTCs as described in the text. When Alice measures the final state
of the $B$ and $A$ qubits to be $|\Phi\rangle$, the standard teleportation
protocol would have Bob (who holds the other half of the initially
entangled state) then holding the state with which Alice measured
the $A$ system. There are two unusual things about this situation:
first, that Bob's system is the same one that Alice wants to teleport,
just at an earlier time, and, second, that Alice can get the outcome
$|\Phi\rangle$ with certainty and so no classical communication to
Bob is required to complete the teleportation.}
\label{fig:p-ctcs}
\end{figure}

The theory of P-CTCs is defined by ignoring the mechanism behind time
travel and postulating only that the result is mathematically equivalent
to teleportation into the past, achieved by the following unphysical
operational protocol schematically illustrated in Fig. \ref{fig:p-ctcs}.

Prepare $2m$ qubits, half labelled $A$ and half $B$, in the maximally
entangled state $|\Phi\rangle=2^{-\frac{m}{2}}\sum_{i}|i\rangle_{B}|i\rangle_{A}$,
where $\{|i\rangle\}_{i}$ is any orthonormal basis for $m$ qubits.
Treat the $B$ qubits as the CV system and let them interact with
the CR qubits as normal. After the interaction, perform a joint projective
measurement on the $B$ and $A$ qubits in a basis that includes $|\Phi\rangle$,
but postselect on the outcome $|\Phi\rangle.$ This is equivalent
to simply projecting the tripartite system of CR, $B$, and $A$ qubits
onto $\langle\Phi|$ and then renormalising the resulting state. Comparing
this to the standard multi-qubit quantum teleportation protocol, the
effect is to ``teleport'' the final state of the $B$ qubits back
onto the $B$ qubits in the past. This protocol may be simulated in
the laboratory by manual postselection of measurement outcomes \cite{Svetlichny09,Lloyd11a}.

The effect of this protocol on CR qubits initially in the pure state
$|\psi_{i}\rangle$ is (up to renormalisation)
\begin{eqnarray}
_{BA}\langle\Phi|U_{\textsc{CR},B}|\psi_{i}\rangle|\Phi\rangle_{BA} & = & 2^{-m}\sum_{i}\,_{B}\langle i|U_{\textsc{CR},B}|i\rangle_{B}|\psi_{i}\rangle\nonumber \\
 & \propto & \Tr_{\textsc{CV}}(U)\,|\psi_{i}\rangle
\end{eqnarray}
where subscripts denote the subsystems to which states belong or upon
which operators act. Generalising this result to mixed input $\rho_{i}$
and renormalising the output to a valid unit-trace density operator
shows that the evolution effected by a P-CTC is
\begin{equation}
\rho_{i}\rightarrow\rho_{f}=\frac{P\rho_{i}P^{\dagger}}{\Tr(P\rho_{i}P^{\dagger})},\quad P\eqdef\Tr_{\textsc{CV}}(U).\label{eq:p-ctcem}
\end{equation}
This evolution is non-linear due to the renormalisation and non-unitary
due to the partial trace. Considered as a map applied to the CR qubits,
Eq. (\ref{eq:p-ctcem}) completely specifies the action of P-CTCs,
and, as such, one may prefer to forget about the protocol described
above and instead define the action of a P-CTC to be given by Eq.
(\ref{eq:p-ctcem}) %
\footnote{This having been said, for specific circuits it is sometimes practically
more convenient to work through the protocol rather than use Eq. (\ref{eq:p-ctcem}).%
}. With this mindset, the protocol demonstrates that P-CTCs are equivalent
to quantum teleportation to the past.

Since Eq. (\ref{eq:p-ctcem}) maps each $\rho_{i}$ onto a specific
$\rho_{f}$ without ambiguity, P-CTCs do not suffer the uniqueness
ambiguity present classically and in D-CTCs and so are not vulnerable
to information paradoxes. Applying the P-CTC protocol to the unproven
theorem circuit described in Sec. \ref{sub:Time-Travel-Paradoxes}
produces pure output $|\psi_{f}\rangle=\frac{1}{\sqrt{2}}\left(|0\rangle_{B}|0\rangle_{M}+|1\rangle_{B}|1\rangle_{M}\right)$.
The output of $N$ such circuits will always be an equal superposition
of all possible theorems and the paradox is avoided.

The operator $P$ defined in Eq. (\ref{eq:p-ctcem}) is not a unitary
operator, but the partial trace of a unitary operator. It does not,
therefore, preserve arbitrary vector norms, but vector norms are still
bounded from above. Consider $P$ acting on a vector $|\psi\rangle$
of the CR system, and let $\{|\alpha\rangle\}_{\alpha=0}^{d-1}$ be
an orthonormal basis on the $d$-dimensional CV system. Using the
triangle inequality and unitarity it is found that 
\begin{eqnarray}
\left\Vert P|\psi\rangle\right\Vert  & \leq & \sum_{\alpha}\left\Vert \langle\alpha|U|\psi\rangle|\alpha\rangle\right\Vert \nonumber \\
 & \leq & \sum_{\alpha}\left\Vert U|\psi\rangle|\alpha\rangle\right\Vert =d\left\Vert \psi\right\Vert .\label{eq:Pbound}
\end{eqnarray}

\subsubsection{Dynamical Consistency and Noise\label{sub:Dynamical-Consistency-and}}

P-CTCs suffer from dynamical consistency paradoxes. The unphysical
protocol described above necessarily fails when there is no $|\Phi\rangle$
component in the state on which to project, \emph{viz.} when Alice
has zero probability of obtaining $|\Phi\rangle$ as her measurement
outcome. In this case $P\rho_{i}P^{\dagger}=0$ and no output state
is defined.

Most authors interpret this to mean that such evolutions ``{[}do{]}
not happen'' \cite{Lloyd11a,Lloyd11b,Brun12}. In terms of the discussion
in Sec. \ref{sub:Time-Travel-Paradoxes}, this is avoiding paradox
by enforcing retrospective constraints to prevent inputs that would
produce a paradox. However, there are also P-CTC circuits where all
inputs lead to a dynamical consistency paradox. One simple, if contrived,
example is to have $U=V\otimes W$, where $W$ is traceless, since
then $P=V\Tr W=0$ regardless of the input state.

If one considers some arbitrary noise on the CV system then dynamical
inconsistency is avoided. So whilst the paradoxes are in the theory,
one would never expect a realistic system to encounter one. The reason
for this is that noise will always introduce an arbitrarily small
component to the state which is consistent. Consider the tripartite
state in the operational protocol before postselection $\sigma=U(\rho_{i}\otimes|\Phi\rangle\langle\Phi|)U^{\dagger}$
and consider some noise channel $\mathsf{N}$ which acts as $\mathsf{N}(\sigma)=(\sigma+\epsilon\chi)/(1+\epsilon)$,
where $\chi$ is some density operator (which may generally depend
on $\sigma$) satisfying $\langle\Phi|\chi|\Phi\rangle\neq0$ and
$\epsilon>0$. If $\mathsf{N}$ acts before the postselection, then
it is seen that if $\langle\Phi|\sigma|\Phi\rangle=0$ then the output
state after postselection is 
\begin{equation}
\rho_{f}=\frac{\langle\Phi|\chi|\Phi\rangle}{\Tr\langle\Phi|\chi|\Phi\rangle}
\end{equation}
independent of $\epsilon$. Any arbitrarily small random noise will
therefore ensure dynamical consistency of P-CTCs, since noise of this
type can never be totally eliminated.

\subsubsection{Physicality\label{sub:Physicality}}

The P-CTC formalism as described does not address the mechanism of
time travel. Moreover the protocol described above to calculate the
effect of P-CTC circuits is manifestly unphysical. Svetlichny described
the formalism as only ``effective quantum time travel'' \cite{Svetlichny09}
that one might achieve in the laboratory by manual postselection,
while Lloyd \emph{et al.} motivated the approach stating that a CTC
is a communication channel into the past and that quantum teleportation
represents a quantum communication channel which can be made to communicate
to the past as described above \cite{Lloyd11b}.

Despite the unphysicality of the approach, the introduction of noise
in Sec. \ref{sub:Dynamical-Consistency-and} is consistent with the
assumption that P-CTC theory produces real effects from real states
and unitaries. In the same way as for D-CTC circuits, noise may be
introduced to P-CTC circuits by modifying the states and unitaries
used accordingly. Inserting such modified elements is equivalent to
modifying them in the P-CTC protocol, even if the protocol is merely
a calculational tool.

The path integral approach of Politzer does give some physical motivation
to the theory of P-CTCs \cite{Politzer94}. Path integrals form an
alternative to quantum circuits as a general approach to theories
of quantum mechanics along CTCs \cite{Echeverria91,Birrell82,Hartle94,Politzer94}.
In Ref. \cite{Lloyd11b} it is shown that P-CTC dynamics is compatible
with the Politzer path integral approach. Neither standard P-CTCs
nor Politzer path integrals assign states to particles along the CTC.
For certain mechanisms of time travel this may not be problematic,
but it would be puzzling if such a state were not defined for a CTC.

One could argue that the theory of P-CTCs has a similar dynamical
ambiguity to D-CTCs since Eq. (\ref{eq:p-ctcem}) defines a physical
change to the state and the point at which this change is made is
not defined in the theory. Just as with D-CTCs, this dynamical ambiguity
has no observable consequences. Also, similarly to the case with D-CTCs,
there is an ambiguity about how to treat mixed states in this non-linear
theory, which is addressed in Sec. \ref{sub:Mixed-states-and}.

\subsubsection{Computation\label{sub:PCTCComputation}}

P-CTCs are able to solve any problem in $\mathsf{PP}$ in polynomial
time \cite{Lloyd11b}, meaning that P-CTCs are thought to be more
powerful than standard $\mathsf{BQP}$ quantum computation. This is
because quantum mechanics with P-CTCs is computationally equivalent
to quantum mechanics with postselection \cite{Brun12} and quantum
mechanics with postselection has the computational power of $\mathsf{PP}$
\cite{Aaronson05}. $\mathsf{PP}$ is contained within $\mathsf{PSPACE}$
and is thought to be strictly smaller \cite{Arora09}. Therefore,
D-CTCs are at least as computationally powerful as P-CTCs, and thought
to be strictly more powerful.

Unlike D-CTCs, P-CTCs are unable to produce discontinuous evolutions
because of the more mild form of non-linearity exhibited, as shown
in Sec. \ref{sub:Non-linearity}.

\subsection{Mixed states and Non-linearity\label{sub:Mixed-states-and}}

\subsubsection{Non-linearity\label{sub:Non-linearity}}

The most general possible form of non-linearity in an equation of
motion is due to the input state appearing at quadratic or higher
orders. D-CTCs are an example of this: Equation (\ref{eq:d-ctcem})
depends on $\rho_{i}$ directly, but also indirectly since $\tau$
depends on $\rho_{i}$.

A special case of non-linearity is \emph{renormalisation non-linearity},
where the equation of motion is linear except for a linearly-determined
scalar factor which simply normalises the final state. P-CTCs are
examples, since Eq. (\ref{eq:p-ctcem}) would be entirely linear were
it not for the denominator, which is simply a normalisation factor.
A non-linear equation may be called \emph{polynomial non-linear} if
it is not renormalisation non-linear.

One way in which renormalisation non-linear equations are more mild
than polynomial non-linear equations is that the former are not able
to lead to discontinuous state evolutions. Consider some input $\rho_{i}$
from some vector space $V$ and some linear equation of motion $\mathsf{M}(\rho_{i})\in V$.
The output is required to be normalised $\mathcal{N}(\rho_{f})=1$
for some linear function $\mathcal{N}:V\rightarrow\mathbb{R}$ and
so we have
\begin{equation}
\rho_{f}(\rho_{i})=\frac{\mathsf{M}(\rho_{i})}{\mathcal{N}(\mathsf{M}(\rho_{i}))}\in V.
\end{equation}
This is a general form for a renormalisation non-linear equation,
and it is simple to see that the state evolution is always continuous
since for any $\sigma\in V$
\begin{eqnarray}
\lim_{\epsilon\rightarrow0}\rho_{f}(\rho_{i}+\epsilon\sigma) & = & \rho_{f}(\rho_{i}).
\end{eqnarray}
Compare this to D-CTCs, which are able to produce non-linear state
evolutions.

\subsubsection{Types of Mixed State\label{sub:Types-of-Mixed}}

In quantum theory one normally identifies two possible types of mixed
state: \emph{proper mixtures} as a result of ignorance of an observer
regarding the ontic state of a system and \emph{improper mixtures}
as a result of only having access to part of an entangled system.
Both are described using density operators with those for improper
mixtures often called reduced density operators. Mixtures giving rise
to the same density operator act identically in all respects regardless
of type; they differ only in preparation procedure and interpretation.

This equivalence is instrumental in enabling quantum theory to have
multiple consistent interpretations. For example, in Everettian quantum
theory there are typically no proper mixtures, only improper mixtures
\footnote{Of course, any Everettian observer could still choose to be ignorant,
but this does not give rise to mixed states in the same way. To clarify,
consider how proper mixtures arise in objective collapse models. An
observer $\mathcal{O}$ sets up a measurement $\mathcal{M}$ on a
system $\mathcal{S}$, but chooses (by not looking, or what have you)
to remain ignorant of the outcome of $\mathcal{M}$. In an objective
collapse model, $\mathcal{O}$ now knows (assuming sufficient understanding
of $\mathcal{S}$, $\mathcal{M}$, and quantum theory) that the state
of the universe has now collapsed into one of multiple possible states
with corresponding probabilities; this ensemble forms a proper mixture
which $\mathcal{O}$ uses to describe $\mathcal{S}$ after $\mathcal{M}$.
In an Everettian model, however, $\mathcal{O}$ now knows that the
universal state has evolved into some macroscopic superposition, the
only uncertainty is about which branch of this $\mathcal{O}$ will
experience upon discovering the outcome of $\mathcal{M}$. For more
on this see, for example, Ref. \cite[especially \S3.2]{Albert10}.%
}, whereas an objective collapse interpretation would allow both types
of mixture.

In a theory with non-linearity, however, it is not valid to describe
proper mixtures using density operators. This follows immediately
because, for a non-linear equation of motion $\mathsf{M(\cdot)}$
acting on some ensemble of states and corresponding probabilities
$\{(\rho_{j},p_{j})\}_{j}$, applying $\mathsf{M}$ to the initial
density operator is not generally the same as the density operator
obtained by applying $\mathsf{M}$ to each individual state in the
ensemble:

\begin{equation}
\mathsf{M}(\sum_{j}p_{j}\rho_{j})\neq\sum_{j}p_{j}\mathsf{M}(\rho_{j}).
\end{equation}
On the other hand, density operators are the correct way to describe
improper mixtures under non-linear evolution. By examining the derivation
of reduced density operators, as given in \cite{Nielsen00} for example,
it is easily seen that linearity of operations is not assumed at any
point.

There is a third type of mixed state, identified in Refs. \cite{Deutsch91,Brun12},
that is rarely discussed but must be taken into account when discussing
time travel. Normally, one assumes that the primitive states of quantum
theory are pure states, implying that the state of the entire universe
is pure, a claim that is mathematically justified due to the purification
theorem \cite{Nielsen00}. One can relax this assumption and allow
mixed states to be primitive states; such mixtures are called \emph{true
mixtures}: mixtures that are not entangled to any reference system
and yet would still be described as mixed by an observer with maximum
knowledge. Ontologically true mixtures are therefore distinct from
both proper and improper mixtures. Since D-CTCs are capable of sending
pure states to mixed states (see, for example, Sec. \ref{sub:Non-uniqueness-Ambiguity-in}),
mixtures arising in this way must be true mixtures, which inevitably
arise in a theory with D-CTCs. Moreover, the state $\tau$ defined
by a D-CTC circuit cannot generally be purified and must be thought
of as a true density operator \cite{Pati10}.

Because reduced density operators are still valid ways of describing
improper mixtures when non-linear evolutions are possible, it follows
that non-linear equations do not treat true and improper mixtures
differently. This includes all theories of time travel presented in
this paper. So whilst theories of time travel may introduce true mixtures
conceptually, they do not affect the way in which calculations are
performed. As such the purification theorem still holds for $\rho_{i}$
and one may always assume that $\rho_{i}=|\psi_{i}\rangle\langle\psi_{i}|$
by simply extending $U$ to act on the purification ancilla as the
identity.

\subsubsection{Non-linearity and Information Processing\label{sub:Non-linearity-and-Information}}

Perhaps the most contentious ambiguity related to theories of quantum
mechanical time travel is how non-linearity affects information processing.

In Ref. \cite{Bennett09} it is argued that D-CTC devices that attempt
to implement information processing tasks on pure state inputs are
impotent in realistic situations. The argument supporting this is
that realistic input to a useful device is always uncertain: Any of
a range of possible inputs could be chosen. It is argued that any
useful device, therefore, acts on mixed state inputs. However, if
the device is non-linear then the desired processing, proven to work
on pure state inputs, does not simply generalise to the realistic
mixed input and so the device will not behave as advertised. This
is called the ``linearity trap'' \cite{Bennett09}.

To resolve this tension, careful attention needs to be paid to assumptions
regarding the ontology of states \cite{Ralph10,Cavalcanti10,Cavalcanti12,Brun12,Pienaar13b}.
In ordinary quantum mechanics one does not generally need to make
such ontological assumptions explicit in order to calculate correctly;
however, in a non-linear theory proper and improper mixtures are no
longer operationally equivalent and therefore ontology becomes important.

A guiding principle may be stated as follows.
\begin{quote}
In any extension of quantum theory including non-linearity, the dynamics
of a system $S$ may generally depend on more than its quantum mechanical
state. One may need to take into account the entire ontic system of
which $S$ is a part and also the manner in which it was prepared
in order to consistently deduce its dynamics.
\end{quote}
In particular, one should be explicit as to when a mixed state is
a proper mixture.

So what of the information processing abilities of D-CTCs and other
non-linear theories? If the ``realistic'' mixed input to such a
device is ontologically a proper mixture, then the action on pure
state inputs should generalise to the mixed state input as the device
is acting on ontological states and is unaware of any particular observer's
ignorance, so the device behaves as advertised. On the other hand,
if the mixed input is improper, then, as pointed out in Ref. \cite{Bennett09},
one cannot generalise in this way.

To illustrate, suppose that the input to some non-linear information
processing device is selected by performing some projective quantum
measurement on a separate system. In Everettian quantum theory the
resulting mixed state is a macroscopic superposition of which only
one branch is experienced, and so the input to the machine is ontologically
an improper mixture. On the other hand, in quantum theory with objective
state collapse the resulting input is a proper mixture. The behaviour
the device would therefore be different depending on the ontology.

\subsubsection{Signalling\label{sub:Signalling}}

Since all non-linear evolutions treat proper and improper mixtures
differently, then so long as this difference is observable it will
lead to an \emph{entanglement detector}: a device capable of telling
whether or not a system is entangled with another. In any interpretation
that involves instantaneous disentanglement by measurement then an
entanglement detector necessarily allows superluminal signalling \cite[\S3.1.3]{Pienaar13b}.
D-CTCs and P-CTCs therefore lead to superluminal signalling with such
an interpretation.

Instantaneous disentanglement is not a necessary feature of quantum
measurement, however. There are many interpretations that do not require
instantaneous disentanglement. Moreover, one can construct alternative
models of quantum measurement that prevent superluminal signalling
in non-linear theories by construction without subscribing to a particular
interpretation \cite{Czachor02,Kent05}. So whilst non-linearity does
not necessarily lead to signalling, it may do depending on the ontology
\cite[\S3]{Pienaar13b}.

As well as this, P-CTCs seem to allow another form of non-locality
due to dynamical consistency paradoxes. If one accepts that evolutions
leading to dynamical consistency paradoxes are disallowed, then this
is to say that the future existence of a P-CTC will affect measurements
before that P-CTC comes into existence in order to avoid the paradox
\cite{Brun12}. However, this type of non-locality is not observable
as dynamical consistency paradoxes with P-CTCs disappear as soon as
noise is added.

\subsubsection{Distinguishing Non-orthogonal States}

Both D-CTCs and P-CTCs are capable of distinguishing non-orthogonal
pure states using a single measurement \cite{Brun09,Brun12}, an impossibility
in standard quantum theory \cite{Nielsen00}. This is neatly illustrated
by considering the single-qubit \emph{distinguishing circuit}: a standard
form circuit with $m=n=1$ and $U=\textsc{CH}_{\textsc{CR},\textsc{CV}}\,\textsc{SWAP}$,
where $\textsc{CH}_{\textsc{CR},\textsc{CV}}$ is the controlled-Hadamard
gate controlled on the CR qubit \cite{Brun09,Brun12}.

Implementing the distinguishing circuit with a D-CTC, it is simple
to check that $\rho_{i}=|0\rangle\langle0|\Rightarrow\tau=|0\rangle\langle0|\Rightarrow\rho_{f}=|0\rangle\langle0|$
and $\rho_{i}=|-\rangle\langle-|\Rightarrow\tau=|1\rangle\langle1|\Rightarrow\rho_{f}=|1\rangle\langle1|$.
This D-CTC therefore renders the non-orthogonal states $|0\rangle$
and $|-\rangle$ distinguishable by a computational basis measurement.

Implementing the same distinguishing circuit with a P-CTC one can
easily verify that the P-CTC operator $P$ of Eq. (\ref{eq:p-ctcem})
is $P=\Tr_{\textsc{CV}}U=|0\rangle\langle0|+|1\rangle\langle-|$.
Therefore, this P-CTC maps $|+\rangle\rightarrow|0\rangle$ and $|1\rangle\rightarrow|1\rangle$,
rendering the non-orthogonal states $|+\rangle$ and $|1\rangle$
distinguishable.

References \cite{Brun09,Brun12} give general recipes for constructing
maps for D-CTC and P-CTC circuits respectively that will map sets
of non-orthogonal states onto sets of orthogonal states, so that the
original states can be distinguished by a single measurement. In this
way, D-CTCs can distinguish non-orthogonal states from any specified
finite set of pure states, allowing D-CTCs to violate the Holevo bound
of classical communication over a quantum channel \cite{Holevo73,Brun09}.
Since P-CTCs are only renormalisation non-linear, they are limited
to only distinguishing pure states from a linearly independent set
and therefore cannot break the Holevo bound in this way.

\subsubsection{Cloning and Deleting}

Distinguishability of non-orthogonal states from an arbitrarily large
set enables arbitrarily precise identification of any state. Therefore,
cloning with D-CTCs is possible with arbitrarily high fidelity by
simply identifying a state and then preparing new copies of it. More
practical approaches to cloning using D-CTCs are presented in Refs.
\cite{Ahn12,Brun13}. However neither cloning method produces clones
with the same entanglement correlations as the original, which is
forbidden by monogamy of entanglement \cite{Coffman00}.

It is possible to use a P-CTC circuit with a single CV qubit to perform
any postselected quantum measurement \cite{Brun12}. It is for this
reason that P-CTCs have the computational power of $\mathsf{PP}$,
as discussed in Sec. \ref{sub:PCTCComputation}. This also means that
P-CTCs trivially violate the no-deleting theorem \cite{Pati00}, which
forbids any process that takes two copies of an unknown pure quantum
state and results in a single copy of the same state with the second
system in some standard ``blank'' state. By simply performing a
postselected measurement, a P-CTC could project the second system
onto the ``blank'' state, a process that would work even if the
original state was orthogonal to the blank state by virtue of any
small noise, as discussed in Sec. \ref{sub:Dynamical-Consistency-and}.

The processes of copying and deleting are dual to one another, and
the impossibility of each in ordinary quantum theory follows due to
linearity.

\section{New Theories\label{sec:New-Theories}}

Having developed the theories of D-CTCs and P-CTCs, the question now
arises as to what other theories of quantum mechanics with time travel
there might be and how they might compare to these existing examples.
Before developing some new theories, it is useful to first review
some background on integrating over quantum states.

\subsection{Integrals over Quantum States}

\subsubsection{Pure States}

Given a quantum system, pure states of that system are described as
unit vectors $|\phi\rangle$ in a Hilbert space $\mathcal{H}$, such
that unit vectors equal up to a phase factor are considered equivalent:
$|\phi\rangle\sim e^{i\theta}|\phi\rangle$. For finite $\dim\mathcal{H}=d$,
the projective space $\mathcal{P}(\mathcal{H})\eqdef\mathbb{C}\mathbf{P}^{d-1}$
contains all of the distinct pure states of that system \cite{Bengtsson06}.

For some scalar function $\mathcal{I}:\mathcal{H}\rightarrow\mathbb{C}$
consider the integral over the physical states

\begin{equation}
I=\int_{\mathcal{P}(\mathcal{H})}\mathrm{d}[\phi]\,\mathcal{I}(\phi),
\end{equation}
where the integration measure $\mathrm{d}[\phi]$ is yet to be defined.
Conveniently, there exists a unique natural measure over $\mathcal{P}(\mathcal{H})$
invariant under unitary transformations given by taking a random unitary
matrix distributed according to the Haar measure on the group $U(d)$
\cite{Zyczkowski01}. Such a measure is usefully written in the Hurwitz
parametrisation \cite{Bengtsson06,Zyczkowski01,Hurwitz97} defined
with respect to an orthonormal basis $\{|\alpha\rangle\}_{\alpha=0}^{d-1}$
on $\mathcal{H}$ such that any pure state $|\phi\rangle$ may be
uniquely written in the form
\begin{multline}
|\phi\rangle=\prod_{\beta=d-1}^{1}\sin\theta_{\beta}|0\rangle\\
+\sum_{\alpha=1}^{d-2}e^{i\varphi_{\alpha}}\cos\theta_{\alpha}\prod_{\beta=d-1}^{\alpha+1}\sin\theta_{\beta}|\alpha\rangle\\
+e^{i\varphi_{d-1}}\cos\theta_{d-1}|d-1\rangle,\label{eq:hurwitz}
\end{multline}
with parameters $\theta_{\alpha}\in[0,\pi/2]$ and $\varphi_{\alpha}\in[0,2\pi)$.
In this parametrisation, the integration measure takes the form
\begin{equation}
\mathrm{d}[\phi(\theta_{\alpha},\varphi_{\alpha})]=\prod_{\alpha=1}^{d-1}\cos\theta_{\alpha}(\sin\theta_{\alpha})^{2\alpha-1}\mathrm{d}\theta_{\alpha}\mathrm{d}\varphi_{\alpha}.\label{eq:integration-measure}
\end{equation}
Equation (\ref{eq:integration-measure}) is unique up to a multiplicative
constant.

As noted above, this natural measure is invariant under unitary operations,
so that under $|\phi\rangle\rightarrow U|\phi\rangle$ the measure
transforms as $\mathrm{d}[\phi]\rightarrow\mathrm{d}[U\phi]=\mathrm{d}[\phi]$.
It may be useful to observe that the Hurwitz parametrisation is a
generalisation of the Bloch sphere parametrisation often used for
qubits $\mathcal{H}=\mathbb{C}^{2}$. For qubits the measure is, up
to a scalar, the rotationally-invariant area measure on a sphere:
\begin{eqnarray}
|\phi\rangle & = & \sin\theta|0\rangle+e^{i\varphi}\cos\theta|1\rangle,\\
\mathrm{d}[\phi] & \propto & \sin(2\theta)\mathrm{d}(2\theta)\mathrm{d}\varphi.
\end{eqnarray}

\subsubsection{Mixed States}

The mixed states of a quantum system with Hilbert space $\mathcal{H}$
and $d=\dim\mathcal{H}$ are described by density operators on $\mathcal{H}$
that are Hermitian, unit trace, and positive semi-definite. The space
of valid density operators on $\mathcal{H}$ is called $\mathcal{D}(\mathcal{H})$
and each operator $\rho\in\mathcal{D}(\mathcal{H})$ describes a distinct
state.

For some scalar valued function $\mathcal{I}:\mathcal{D}(\mathcal{H})\rightarrow\mathbb{C}$
consider the integral over the density operators on $\mathcal{H}$,
\begin{equation}
I=\int_{\mathcal{D}(\mathcal{H})}\mathrm{d}[\tau]\,\mathcal{I}(\tau),
\end{equation}
for some integration measure $\mathrm{d}[\tau]$. Unlike $\mathbb{C}\mathbf{P}^{d-1}$,
there is no unique natural measure on $\mathcal{D}(\mathcal{H})$
\cite{Bengtsson06} and so one has to be chosen, along with a useful
way to parametrise $\tau$. As a result, there is no unique natural
way to define $I$; it will depend on the choice of measure used.

\subsection{Desirable Features\label{sub:Desirable-Features}}

When considering how a new theory of quantum mechanics with time travel
might be developed, it is useful to consider how it might be desirable
for such a theory to behave. Having already developed D-CTCs and P-CTCs
fully it is easier to anticipate potential shortcomings of any new
theory. A list of desirable features is given below. Of course, all
desiderata are linked to various philosophical prejudices, but there
is still utility in considering them.
\begin{enumerate}
\item The theory should have physical motivation and have a physical interpretation.
\item The theory should reproduce standard quantum mechanics well enough
to be consistent with current observations. In the case of CTCs it
should reproduce quantum mechanics locally along the CTC, as well
as in spacetime regions far from the CTC. It is also expected to be
locally approximately consistent with special relativity and relativistic
causal structure. Specifically, it should not allow superluminal signalling.
\item The theory should be dynamically consistent for all choices of $U$
and $\rho_{i}$. In other words, it should not have disallowed evolutions
that lead to dynamical consistency paradoxes.
\item The theory should specify $\rho_{f}$ uniquely given $U$ and $\rho_{i}$.
If multiple possible output states are considered, then probabilities
for each of these should be specified. In other words, it should not
have uniqueness ambiguities that lead to information paradoxes (Sec.
\ref{sub:Time-Travel-Paradoxes}).
\item The theory should specify a state $\tau$ that travels back in time;
this should either be uniquely specified or an ensemble of possibilities
with corresponding probabilities should be uniquely specified.
\item Given a pure $\rho_{i}$, prejudice might require either or both of
$\rho_{f}$ and $\tau$ to also be pure.
\item The theory should not be able to distinguish non-orthogonal states
in a single measurement, neither should it be able to clone arbitrary
quantum states.
\end{enumerate}
Feature 1 is the most subtle of these, and is discussed for D-CTCs
and P-CTCs in Secs. \ref{sub:Interpretation-of-D-CTCs} and \ref{sub:Physicality}
respectively.

D-CTCs have feature 2 so long as ontological assumptions regarding
collapse are made that rule out superluminal signalling. P-CTCs only
have feature 2 in the presence of finite noise, and even then similar
assumptions about collapse are required to rule out signalling. However,
as noted in Sec. \ref{sub:Signalling}, adding any non-linear evolution
to quantum mechanics opens up the possibility of signalling in this
way.

Feature 3 is owned by D-CTCs but not P-CTCs, while feature 4 is definitely
owned by P-CTCs but is only owned by D-CTCs by adding an extra postulate
(uniqueness with D-CTCs may also be gained by finite noise, as conjectured
in Sec. \ref{sub:Non-uniqueness-Ambiguity-in}).

Neither the theory of D-CTCs nor P-CTCs fully has feature 5. P-CTCs
do not specify any $\tau$, while D-CTCs specify $\tau$ but not necessarily
uniquely. Feature 6 is perhaps the least compelling feature listed
and is one that neither D-CTCs nor P-CTCs have. Feature 7 is also
not one respected by either P-CTCs or D-CTCs.

It is a lot to ask for a theory to satisfy these desiderata. Notably,
the standard way of introducing time travel into classical mechanics
does not satisfy 2, 3, 4, or 5. However, since it is not clear how
to proceed with constructing its quantum analogue, such a list may
be a helpful guide.

\subsection{A Selection of New Theories\label{sub:A-Selection-of}}

With the above desiderata in mind, new theories of quantum mechanics
with time travel may be constructed. In this section two overlapping
classes of new theories are considered:\emph{ weighted D-CTCs} and
\emph{transition probability theories}. An example of the latter,
dubbed \emph{T-CTCs}, is selected for further study.

Weighted D-CTCs represent an extension of the theory of D-CTCs. These
are described by parametrising the convex subset of density operators
$\tau_{\alpha}$ allowed by the consistency condition (\ref{eq:d-ctcem})
with $\alpha$, and then assigning a weight $w_{\alpha}\geq0$ to
each. The weighted mixture of these is then used for $\tau=\int\mathrm{d}\alpha\, w_{\alpha}\tau_{\alpha}/\int\mathrm{d}\alpha\, w_{\alpha}$.
The D-CTC protocol can then be used with this uniquely determined
choice of $\tau$. One basic example would be to weight all possibilities
equally $w_{\alpha}=1$, giving a \emph{uniform} weighted D-CTC theory.
There is a whole class of theories on this theme, from different choices
of the weights to only taking the mixture over a subset of the $\tau_{\alpha}$
(perhaps taking a mixture of the $\tau_{\alpha}$ with the minimum
entropy). In terms of the desirable features listed, this theory would
gain features 4 and 5 at least, possibly at the expense of feature
1 depending on the details and motivation of the theory. Such a theory
is essentially that of D-CTCs, with an alternative to the maximum
entropy rule.

Transition probability theories make use of some useful intuition
from standard quantum mechanics. It is common to say that the probability
of an initial state $|I\rangle$ to transition into a final state
$|F\rangle$ under the unitary transformation $V$ is given by the
transition probability $|\langle F|V|I\rangle|^{2}$. More precisely,
what is meant is that $|\langle F|V|I\rangle|^{2}$ is the Born rule
probability of finding the system in state $|F\rangle$ if one were
to measure the system to see if it were in state $|F\rangle$ after
the transformation. As an example of this useful way of thinking consider
starting with a bipartite system, initially in state $|\psi_{i}\rangle|\phi\rangle$,
and act upon it with the unitary $U$; the ``probability of finding
the second system in $|\phi\rangle$'' after the transformation is
$p(\phi)=\left\Vert \langle\phi|U|\psi_{i}\rangle|\phi\rangle\right\Vert ^{2}$
\footnote{One needs to be careful with expressions such as these since different
states belong on different Hilbert sub-spaces, while $U$ acts on
the entire Hilbert space. Throughout this paper, $|\phi\rangle$ is
used for a pure state on the CV Hilbert space and $\tau$ used for
a mixed state on the same space, while $|\psi_{i,f}\rangle$ and $\rho_{i,f}$
are both used for pure and mixed states respectively on the CR Hilbert
space. The various identity operators and tensor products will be
left implicit. It is hoped that this will avoid confusion as to how
such expressions fit together.%
}. 

There is no unique way to generalise the transition probability to
mixed states $\rho$ and $\tau$. One option is to use the Hilbert-Schmidt
inner product $\Tr[\rho\tau]$, which is the probability for $\rho$
to be found in an eigenstate of the proper mixture $\tau$ and for
$\tau$ to be a realisation of that same eigenstate, averaged over
all eigenstates of $\tau$. Another way to think about this is that
if $\rho$ is measured using some POVM which has an element $\tau$,
then the probability of the corresponding result being obtained is
$\Tr[\rho\tau]$ \cite{Nielsen00}. Another option is the square of
the fidelity \cite{Uhlmann11}, Eq. (\ref{eq:fidelity}), the interpretation
of which involves considering $\rho$ as an improper mixture on which
a measurement is performed by projective measurement of the larger
purified system. Both of these options reduce to the transition probability
in the case of both states being pure. The use of any mixed state
transition probability must be motivated by its operational meaning
in context. For the remainder of this paper it will be assumed that
the appropriate generalisation of transition probability to mixed
states is the Hilbert-Schmidt inner product, so that the probability
for a bipartite system initially in the state $\rho_{i}\otimes\tau$
to have the second system found in the state $\tau$ after some unitary
transformation $U$ is given by $p(\tau)=\Tr\left[\tau U(\rho_{i}\otimes\tau)U^{\dagger}\right]$
\footnote{One curiosity of using this generalisation of transition probability
is that the probability for $\rho$ to transition to $\rho$ under
unitary $\mathbbm{1}$ is strictly less than unity for mixed $\rho$.
This is simply a reflection of the fact that mixed states can be viewed
as epistemic states over the pure states.%
}.

The transition probability theories are the specific theories obtained
by applying these ideas to time travel. The choices that need to be
made to be define a specific theory include: whether pure or mixed
CV states are used, which $|\phi\rangle$ or $\tau$ are to be considered,
and how $\rho_{f}$ should be separated from the CV system?

The following is an example of a transition probability theory that
is also a weighted D-CTC theory. That is, choose the weights of a
weighted D-CTC theory to be the transition probabilities: $w_{\alpha}=p(\tau_{\alpha})=\Tr[\tau_{\alpha}^{2}]$.
Therefore, the equation of motion is 
\begin{equation}
\rho_{f}=\frac{\int\mathrm{d}\alpha\Tr\left[\tau_{\alpha}^{2}\right]\Tr_{\textsc{CV}}\left[U(\rho_{i}\otimes\tau_{\alpha})U^{\dagger}\right]}{\int\mathrm{d}\alpha\Tr\left[\tau_{\alpha}^{2}\right]}.
\end{equation}

Another collection of transition probability theories is found by
integrating over all possible initial CV states, weighted by the corresponding
transition probability, and so 
\begin{equation}
\tau_{i}=Z^{-1}\int_{\mathcal{P}(\mathcal{H}_{\textsc{CV}})}\mathrm{d}[\phi]\, p(\phi)|\phi\rangle\langle\phi|\label{eq:transitiontaui}
\end{equation}
is used if pure CV states are considered, or similarly with an integral
over $\mathcal{D}(\mathcal{H}_{\textsc{CV}})$ if mixed CV states
care considered. $Z>0$ normalises the state $\tau_{i}$ and $p(\phi)$
is the transition probability as described above. There are two options
for then specifying $\rho_{f}$: Either take the partial trace as
with D-CTCs or use the same partial projection used to calculate the
transition probability. The first of these, using pure CV states,
gives a theory with the equation of motion
\begin{equation}
\rho_{f}=Z^{-1}\int\mathrm{d}[\phi]\, p(\phi)\Tr_{\textsc{CV}}\left[U(|\psi_{i}\rangle\langle\psi_{i}|\otimes|\phi\rangle\langle\phi|)U^{\dagger}\right],\label{eq:pureTICTCs}
\end{equation}
which is one of various theories on this theme.

Another such transition probability theory is the \emph{theory of
T-CTCs}, which is the theory found by integrating over all pure CV
states, weighted by transition probability, but using the partial
projection to find $\rho_{f}$. T-CTCs are developed fully below in
Sec. \ref{sub:T-CTCs}, complete with a discussion of their physical
motivation and interpretation. Some of the other theories that are
variations on this theme are briefly considered in Sec. \ref{sub:Relation-to-Alternative}.

\subsection{The Uniqueness Ambiguity and Epistemic Reasoning\label{sub:The-Uniqueness-Ambiguity}}

Before proceeding to detail the theory of T-CTCs, some remarks are
in order about the uniqueness ambiguity. In Sec. \ref{sub:Time-Travel-Paradoxes}
this ambiguity was introduced as a necessary and sufficient condition
for a theory to suffer information paradoxes, as defined in that section.
The argument hinges on the idea that if the final state is uniquely
determined by the initial state and dynamics, then what has occurred
is regarded as a (possibly very powerful) computation and is therefore
not paradoxical according to that definition.

The argument still holds if the unique final state is an epistemic
state, so long as the probabilities in the epistemic state are determined
by the physics rather than purely epistemic principles. If the probabilities
are physically determined then any new information obtained can be
viewed as being due to a probabilistic computation. For any particular
final state to be likely, the physics must not only establish that
final state as a possibility, but also that the corresponding probability
is sufficiently high. Models for probabilistic computation are well-established
and certainly not paradoxical.

Compare this to D-CTCs without either noise or the maximum entropy
rule. In Sec. \ref{sub:Non-uniqueness-Ambiguity-in} it was claimed
that the theory of D-CTCs contains the uniqueness ambiguity and therefore
suffers from information paradoxes. This is different from a probabilistic
computation since D-CTCs assign no probabilities to the possible final
states; they are merely left as possibilities.

Therefore, as far as the definitions in Sec. \ref{sub:Time-Travel-Paradoxes}
go, information paradoxes are still avoided if uniqueness ambiguity
is avoided; \emph{viz.}, when a unique physically determined epistemic
state is specified. It is for this reason that requirements 4 and
5 of Sec. \ref{sub:Desirable-Features} allow for uniquely specified
epistemic states.

For example, suppose a time travel circuit is designed to produce
previously unknown theorems. If this circuit produces unique theorems
with certainty from the input, then, as discussed in Sec. \ref{sub:Time-Travel-Paradoxes},
this is a type of computation: a novel automated theorem prover. Similarly,
if the circuit produces one of a selection of possible theorems from
the input, each with a given probability, then the circuit is performing
a (possibly novel) probabilistic computation. On the other hand, if
a theory allows for a circuit that could produce one of a range of
possible theorems but has no way of giving the probability of each
obtaining, then that is an information paradox.

\subsection{T-CTCs\label{sub:T-CTCs}}

The theory of T-CTCs may be motivated as follows. Consider a CR observer
watching a standard form time travel circuit evolve and suppose that
the primitive states of quantum theory are pure states.

This observer watches a CV system emerge from the future in some unknown
pure state $|\phi\rangle$. This is then observed to interact with
a CR system, initially in state $|\psi_{i}\rangle$, via unitary $U$.
The CV system then proceeds to head back in time. At this point, the
observer may judge whether any given $|\phi\rangle$ is a consistent
initial CV state. If someone were to have measured the CV system immediately
before it travelled back in time, then the probability of their finding
that any given $|\phi\rangle$ is a consistent initial state is $p(\phi)=\left\Vert \langle\phi|U|\psi_{i}\rangle|\phi\rangle\right\Vert ^{2}$.
So for any pair $|\phi_{1}\rangle$ and $|\phi_{2}\rangle$, the former
would be found to be consistent $p(\phi_{1})/p(\phi_{2})$ times more
often than the latter. It therefore seems reasonable to conclude that
$|\phi_{1}\rangle$ is $p(\phi_{1})/p(\phi_{2})$ times more likely
to have been the initial state than $|\phi_{2}\rangle$. The observer
therefore considers $\tau_{i}$ as a proper mixture over all $|\phi\rangle\in\mathcal{P}(\mathcal{H}_{\textsc{CV}})$,
each weighted by $p(\phi)$, Eq. (\ref{eq:transitiontaui}). On the
other hand, consistency demands that if $|\phi\rangle$ was the initial
CV state, then on heading back in time the CV system must be found
to be in the same state again. So the observer can describe the final
state of the CR system in each case by the partial projection $\langle\phi|U|\psi_{i}\rangle|\phi\rangle/\left\Vert \langle\phi|U|\psi_{i}\rangle|\phi\rangle\right\Vert ^{2}$
consistent with this being the case. The resulting final state for
the CR system is, therefore,

\begin{eqnarray}
\rho_{f} & = & Z^{-1}\int\mathrm{d}[\phi]\, U_{\phi}|\psi_{i}\rangle\langle\psi_{i}|U_{\phi}^{\dagger},\label{eq:TCTCeom}\\
U_{\phi} & = & \langle\phi|U|\phi\rangle,\\
Z & = & \int\mathrm{d}[\phi]\,\langle\psi_{i}|U_{\phi}^{\dagger}U_{\phi}|\psi_{i}\rangle,\label{eq:TCTCnormalisation}
\end{eqnarray}
where the operator $U_{\phi}$ acts only on $\mathcal{H}_{\textsc{CR}}$
and the constant $Z>0$ is defined to normalise $\rho_{f}$.

Similar arguments may be used to motivate other theories, such as
some of the other transition probability theories mentioned in Sec.
\ref{sub:A-Selection-of}. The above is not intended as a derivation,
but a motivational explanation for T-CTCs akin to those given for
D-CTCs and P-CTCs. As with those theories, T-CTCs are defined by the
Eqs. (\ref{eq:TCTCeom}-\ref{eq:TCTCnormalisation}) rather than by
any interpretation.

Several features of the theory immediately follow from the definition
in Eqs. (\ref{eq:TCTCeom}-\ref{eq:TCTCnormalisation}). First, it
is a non-unitary and non-linear theory. Second, it is only renormalisation
non-linear and as such it only gives rise to continuous evolutions.
Third, there is no ambiguity in the equation of motion (\ref{eq:TCTCeom}),
so there is no uniqueness ambiguity and no information paradoxes.
Before proceeding to consider what other features T-CTCs may have,
it is first necessary to show that T-CTCs satisfy some basic consistency
requirements. It is also convenient to re-write Eq. (\ref{eq:TCTCeom})
in a simpler form.

\subsubsection{Basic Requirements\label{sub:Basic-Requirements}}

Any reasonable theory of quantum mechanics with time travel should
satisfy some basic consistency criteria, and it is necessary to show
that T-CTCs satisfy them too before proceeding. Needless to say, both
D-CTCs and P-CTCs satisfy these criteria.

The first is that if the identity is applied to the CV system, then
the equation of motion should just reduce to ordinary quantum unitary
evolution. To see this, suppose $U=V\otimes W$ separates on the CR
and CV systems. Then $U_{\phi}=\langle\phi|W|\phi\rangle V$ and the
CR system factorises out of the integral so that $\rho_{f}=V|\psi_{i}\rangle\langle\psi_{i}|V^{\dagger}$
as required.

The second is to show that the point at which the projection occurs
makes no difference. This is equivalent to saying that there is no
observable dynamical ambiguity in this theory, in the sense discussed
in Secs. \ref{sub:D-CTCs} and \ref{sub:Physicality} for D-CTCs and
P-CTCs respectively. This can be seen by considering the transformations
$U\rightarrow U(\mathbbm{1}\otimes W)$ and $U\rightarrow(\mathbbm{1}\otimes W)U$
for some $W$ that only acts on the CV system. By unitary invariance
of the measure, one sees that both transformations have the same result,
so it does not matter whether one applies $W$ before or after $U$
(equivalently, before or after the transition probabilities were calculated).

\subsubsection{Simplification of the T-CTC Equation of Motion}

The equation of motion (\ref{eq:TCTCeom}) for T-CTCs in its current
form is rather opaque. In order to more easily calculate with the
theory, it is useful to perform the integration in generality and
therefore simplify the equation.

Let $\{|\alpha\rangle\}_{i=0}^{d-1}$ be an orthonormal basis for
the $d$-dimensional CV system and expand the unitary $U$ in the
Kronecker product form in this basis $U=\sum_{\alpha,\beta}A_{\alpha\beta}\otimes|\alpha\rangle\langle\beta|$,
where $A_{\alpha\beta}$ are operators on the CR system. In this form
the equation of motion is 
\begin{equation}
\rho_{f}=Z^{-1}\sum_{\alpha,\beta,\gamma,\delta}I_{(\alpha\beta)(\gamma\delta)}A_{\alpha\beta}|\psi_{i}\rangle\langle\psi_{i}|A_{\gamma\delta}^{\dagger},\label{eq:TCTCsimplification1}
\end{equation}
having defined the integrals
\begin{equation}
I_{(\alpha\beta)(\gamma\delta)}=\int\mathrm{d}[\phi]\,\langle\phi|\alpha\rangle\langle\beta|\phi\rangle\langle\phi|\delta\rangle\langle\gamma|\phi\rangle.\label{eq:TCTCsimplification2}
\end{equation}

Now consider expanding both $\mathrm{d}[\phi]$ and $|\phi\rangle$
in the Hurwitz parametrisation {[}Eqs. (\ref{eq:hurwitz},\ref{eq:integration-measure}){]}
with respect to the same basis. Since, for each $\alpha$, $\mathrm{d}\varphi_{\alpha}$
factorises out of the measure, any integrand in which the only $\varphi_{\alpha}$-dependence
is an integer power of $e^{i\varphi_{\alpha}}$ will integrate to
zero. Considering the integrals in Eq. (\ref{eq:TCTCsimplification2}),
every integrand will have such a phase factor unless one or both of
the following conditions is met: $\alpha=\beta$ and $\gamma=\delta$,
or $\alpha=\gamma$ and $\beta=\delta$. In these cases, all phase
factors will cancel out and the phase integrals will not come to zero.
Discarding the zero integrals in Eq. (\ref{eq:TCTCsimplification1})
it is, therefore, found that
\begin{multline}
\rho_{f}=Z^{-1}\left(\sum_{\alpha\neq\beta}I_{(\alpha\beta)(\alpha\beta)}A_{\alpha\beta}|\psi_{i}\rangle\langle\psi_{i}|A_{\alpha\beta}^{\dagger}\right.\\
+\sum_{\alpha\neq\beta}I_{(\alpha\alpha)(\beta\beta)}A_{\alpha\alpha}|\psi_{i}\rangle\langle\psi_{i}|A_{\beta\beta}^{\dagger}\\
+\left.\sum_{\alpha}I_{(\alpha\alpha)(\alpha\alpha)}A_{\alpha\alpha}|\psi_{i}\rangle\langle\psi_{i}|A_{\alpha\alpha}^{\dagger}\right).\label{eq:TCTCsimplification3}
\end{multline}

By unitary invariance of the integration measure one may rotate $|\phi\rangle$
in each of these integrals so that only the $|d-1\rangle$ and $|d-2\rangle$
components contribute, so that for $\alpha\neq\beta$
\begin{multline}
I_{(\alpha\beta)(\alpha\beta)}=I_{(\alpha\alpha)(\beta\beta)}=\int\mathrm{d}[\phi]\,|\langle\phi|d-1\rangle|^{2}|\langle\phi|d-2\rangle|^{2}\\
=(2\pi)^{d-1}\left(\int\prod_{\gamma=1}^{d-3}\cos\theta_{\gamma}(\sin\theta_{\gamma})^{2\gamma-1}\mathrm{d}\theta_{\gamma}\right)\\
\times\int\mathrm{d}\theta_{d-1}\mathrm{d}\theta_{d-2}\cos^{3}\theta_{d-1}\cos^{3}\theta_{d-2}\sin^{2d-1}\theta_{d-1}\sin^{2d-5}\theta_{d-2}\label{eq:tctcintegral1}
\end{multline}
where in the final line the integrand has been expanded out in the
Hurwitz parametrisation. Similarly 
\begin{multline}
I_{(\alpha\alpha)(\alpha\alpha)}=\int\mathrm{d}[\phi]\,|\langle\phi|d-1\rangle|^{4}\\
=(2\pi)^{d-1}\left(\int\prod_{\gamma=1}^{d-3}\cos\theta_{\gamma}(\sin\theta_{\gamma})^{2\gamma-1}\mathrm{d}\theta_{\gamma}\right)\\
\times\int\mathrm{d}\theta_{d-1}\mathrm{d}\theta_{d-2}\cos^{5}\theta_{d-1}\cos\theta_{d-2}\sin^{2d-3}\theta_{d-1}\sin^{2d-5}\theta_{d-2}.\label{eq:tctcintegral2}
\end{multline}
By evaluating the final lines of Eqs. (\ref{eq:tctcintegral1}) and
(\ref{eq:tctcintegral2}) it is seen that, for $\alpha\neq\beta$,
$I_{(\alpha\alpha)(\alpha\alpha)}/I_{(\alpha\beta)(\alpha\beta)}=2$.
So that from Eq. (\ref{eq:TCTCsimplification3}) one finds
\begin{equation}
\rho_{f}\propto\sum_{\alpha,\beta}\left(A_{\alpha\beta}|\psi_{i}\rangle\langle\psi_{i}|A_{\alpha\beta}^{\dagger}+A_{\alpha\alpha}|\psi_{i}\rangle\langle\psi_{i}|A_{\beta\beta}^{\dagger}\right).
\end{equation}

Finally, note the following identities, which may readily be verified
by expanding the traces: $P\eqdef\Tr_{\textsc{CV}}U=\sum_{\alpha}A_{\alpha\alpha}$
and $\sum_{\alpha,\beta}A_{\alpha\beta}|\psi_{i}\rangle\langle\psi_{i}|A_{\alpha\beta}^{\dagger}=\Tr_{\textsc{CV}}\left[U\left(|\psi_{i}\rangle\langle\psi_{i}|\otimes\mathbbm{1}\right)U^{\dagger}\right]$.
Using these, and introducing a normalising scalar $z>0$ (which is
generally different from $Z$ used before), the final form of the
equation of motion becomes
\begin{equation}
\rho_{f}=z^{-1}\left(P|\psi_{i}\rangle\langle\psi_{i}|P^{\dagger}+d\Tr_{\textsc{CV}}\left[U\left(|\psi_{i}\rangle\langle\psi_{i}|\otimes\frac{\mathbbm{1}}{d}\right)U^{\dagger}\right]\right).\label{eq:TCTCsimpleeom}
\end{equation}

Equation (\ref{eq:TCTCsimpleeom}) is in a much more revealing form
than Eq. (\ref{eq:TCTCeom}). It shows that the T-CTC equation of
motion is a weighted mixture of the corresponding P-CTC equation of
motion (\ref{eq:p-ctcem}) with an ordinary quantum channel. This
can give the impression that T-CTCs are akin to noisy P-CTCs.

\subsubsection{Paradoxes}

The theory of T-CTCs is always dynamically consistent. This may be
seen directly from the equation of motion in the form of Eq. (\ref{eq:TCTCsimpleeom}).
Even though it is possible for $P|\psi_{i}\rangle=0$, the second
term will always give rise to a non-zero density operator. Since it
has already been seen that T-CTCs do not suffer uniqueness ambiguities,
it follows that T-CTCs contain neither type of paradox identified
in Sec. \ref{sub:Time-Travel-Paradoxes}. Unlike P-CTCs and D-CTCs,
no noise or extra rule is required to avoid these paradoxes.

For example, consider applying Eq. (\ref{eq:TCTCsimpleeom}) to the
unproven theorem circuit of Sec. \ref{sub:Time-Travel-Paradoxes}.
For this circuit, 
\begin{multline}
P=|0\rangle_{B}\langle0|\otimes|0\rangle_{M}\langle0|+|0\rangle_{B}\langle1|\otimes|1\rangle_{M}\langle1|\\
+|1\rangle_{B}\langle1|\otimes|0\rangle_{M}\langle1|+|1\rangle_{B}\langle0|\otimes|1\rangle_{M}\langle0|
\end{multline}
and therefore $P|00\rangle_{BM}=|00\rangle_{BM}+|11\rangle_{BM}$.
It is also easily found that $\Tr_{\textsc{CV}}\left[U(|00\rangle_{BM}\langle00|\otimes\mathbbm{1})U^{\dagger}\right]=|00\rangle_{BM}\langle00|+|11\rangle_{BM}\langle11|$.
Substituting these into Eq. (\ref{eq:TCTCsimpleeom}) the output the
unproven theorem T-CTC circuit is, therefore,
\begin{multline}
\rho_{f}=\frac{1}{2}|00\rangle_{BM}\langle00|+\frac{1}{4}|00\rangle_{BM}\langle11|\\
+\frac{1}{4}|11\rangle_{BM}\langle00|+\frac{1}{2}|11\rangle_{BM}\langle11|.
\end{multline}
This same result may also be found, with rather more effort, directly
from the integral expression Eq. (\ref{eq:TCTCeom}).

\subsubsection{Computation}

Basic conclusions on the computational power of T-CTCs follow directly
from Eq. (\ref{eq:TCTCsimpleeom}). The second term is realisable
in ordinary quantum mechanics, and so is limited to the power of $\mathsf{BQP}$,
while the first term is the P-CTC equation of motion. Since $\mathsf{BQP}\subseteq\mathsf{PP}$
it follows that T-CTCs cannot efficiently solve any problems that
are not contained within $\mathsf{PP}$. There may be problems in
$\mathsf{PP}$ that they are not able to solve efficiently, so P-CTCs
are at least as powerful as T-CTCs and may be strictly more powerful.

Moreover, the form of Eq. (\ref{eq:TCTCsimpleeom}) suggests that
T-CTCs may be less powerful than P-CTCs. This is because, for a T-CTC,
$\rho_{f}$ is only a pure state if either $P|\psi_{i}\rangle=0$
or if the two terms in Eq. (\ref{eq:TCTCsimpleeom}) are equal. So
every T-CTC algorithm that outputs a pure state is achievable on an
ordinary quantum computer in exactly the same way. It would require
great cunning to design an algorithm for a T-CTC-equipped computer
that made computational use of the first term in Eq. (\ref{eq:TCTCsimpleeom}).
This observation also prevents T-CTCs from being able to perform an
arbitrary postselected quantum measurement, since many postselected
measurement outcomes are pure states. Therefore, one cannot prove
that T-CTCs have the power of $\mathsf{PP}$ in the same way as seen
for P-CTCs in Sec. \ref{sub:PCTCComputation}.

\subsubsection{Mixed States and Non-linearity\label{sub:Mixed-States-and}}

In Sec. \ref{sub:Mixed-states-and} it was noted that both improper
and true mixtures are validly described with density operators in
non-linear extensions of quantum theory, and that the purification
still holds so that $\rho_{i}=|\psi_{i}\rangle\langle\psi_{i}|$ may
always be assumed. This remains true in the theory of T-CTCs. It also
remains true that proper mixtures are not validly described by density
operators and that non-linearity of T-CTCs opens the possibility of
creating an entanglement detector and therefore the possibility of
signalling exactly as with D-CTCs and P-CTCs.

Another consequence of non-linearity is that both D-CTCs and P-CTCs
are capable of distinguishing non-orthogonal states in a single measurement.
However, it is now seen that this is not the case with T-CTCs. 

Consider the problem of distinguishing between two states $\rho$
and $\sigma$. The probability of success when using a single optimal
measurement is given by $\frac{1}{2}[1+D(\rho,\sigma)]$, where $D(\rho,\sigma)\eqdef\frac{1}{2}\Tr|\rho-\sigma|$
is the trace distance between the states \cite{Fuchs99}. Therefore,
$\rho$ and $\sigma$ are perfectly distinguishable in a single measurement
if and only if $D(\rho,\sigma)=1$.

Another measure of distinguishability of states is the the fidelity
between $\rho$ and $\sigma$, defined as \cite{Nielsen00}
\begin{equation}
F(\rho,\sigma)\eqdef\Tr\sqrt{\rho^{1/2}\sigma\rho^{1/2}}.\label{eq:fidelity}
\end{equation}
In the case of pure states $|a\rangle$ and $|b\rangle$, the Fidelity
takes on the particularly simple form $F(|a\rangle,|b\rangle)=|\langle a|b\rangle|$. 

So, suppose one wishes to distinguish quantum states using a T-CTC.
Only pure state inputs need be considered, so what is required is
a bound on the distinguishability of the output states of some T-CTC
circuit, $\rho_{f}^{a}$ and $\rho_{f}^{b}$, for which the input
states were $|a\rangle$ and $|b\rangle$, respectively. Note first
that trace distance is bounded by fidelity \cite{Nielsen00}
\begin{equation}
D(\rho_{f}^{a},\rho_{f}^{b})\leq\sqrt{1-F(\rho_{f}^{a},\rho_{f}^{b})^{2}}.\label{eq:disitinguishability1}
\end{equation}
It is then useful to separate the terms of $\rho_{f}^{a,b}$ seen
in Eq. (\ref{eq:TCTCsimpleeom}). Therefore, write $\rho_{f}^{\psi}=(1-\lambda^{\psi})\sigma^{\psi}+\lambda^{\psi}\tau^{\psi}$,
where $\tau^{\psi}=\Tr[U(|\psi\rangle\langle\psi|\otimes\frac{\mathbbm{1}}{d})U^{\dagger}]$
and 
\begin{equation}
\lambda^{\psi}=\frac{d}{d+\left\Vert P|\psi\rangle\right\Vert ^{2}}\geq\frac{1}{d+1}\label{eq:disitinguishability2}
\end{equation}
where the inequality is a result of Eq. (\ref{eq:Pbound}). Using
strong concavity and monotonicity of fidelity under quantum operations
\cite{Nielsen00} it is seen that 
\begin{eqnarray}
F(\rho_{f}^{a},\rho_{f}^{b}) & \geq & \sqrt{\lambda^{a}\lambda^{b}}F(\tau^{a},\tau^{b})\nonumber \\
 & \geq & \sqrt{\lambda^{a}\lambda^{b}}F(|a\rangle,|b\rangle)\geq\frac{1}{d+1}|\langle a|b\rangle|.\label{eq:disitinguishability3}
\end{eqnarray}
Finally, observe that by Eqs. (\ref{eq:disitinguishability1}-\ref{eq:disitinguishability3})
\begin{equation}
D(\rho_{f}^{a},\rho_{f}^{b})\leq\sqrt{1-\frac{|\langle a|b\rangle|^{2}}{(d+1)^{2}}}\leq1,
\end{equation}
with equality to unity only possible if $\langle a|b\rangle=0$.

This proves that the output states of a T-CTC circuit are only perfectly
distinguishable from one another in a single measurement if the input
states were. It leaves open the question as to whether one may use
T-CTCs to probabilistically distinguish non-orthogonal states with
greater success than standard quantum mechanics.

It is comparatively very simple to observe that T-CTCs are incapable
of cloning pure states. A pure state cloning machine always outputs
a pure state. Since any T-CTC outputting a pure state can be simulated
exactly by an ordinary quantum operation, then the no-cloning theorem
for T-CTCs is simply a result of the no-cloning theorem in ordinary
quantum theory. In exactly the same way, it also immediately follows
that T-CTCs are incapable of deleting arbitrary pure states. However,
the question as to whether mixed states can be broadcast \cite{Barnum96}
is left open.

\subsubsection{Relation to Alternative Theories\label{sub:Relation-to-Alternative}}

Several of the above results for T-CTCs are easily modified to apply
to some of the closely related theories introduced in Sec. \ref{sub:A-Selection-of}.

Consider the modification to T-CTCs where, instead of integrating
over pure CV states, mixed CV states are integrated over. This represents
a class of theories since there is no unique natural choice for the
integration measure, but it is still possible to deduce some general
properties. Assuming that the chosen integration measure is unitarily
invariant, this theory satisfies the basic requirements considered
in Sec. \ref{sub:Basic-Requirements}. This theory is also only renormalisation
non-linear and so continuity follows immediately. It is also possible
to show that this theory always defines a unique non-zero $\rho_{f}$
for every $\rho_{i}$ and $U$, so that the theory suffers neither
dynamical consistency nor information paradoxes %
\footnote{The proof that $\rho_{f}$ is non-zero follows by showing that the
integrand is positive semi-definite and that there exist some $\tau$
for which the integrand is non-zero. Importantly, for this to go through
an assumption does need to be made about the positivity of the chosen
measure.\label{fn:The-proof-that}%
}.

Now consider the modification to T-CTCs expressed in Eq. (\ref{eq:pureTICTCs}),
where, instead of separating CV and CR systems by a projection, they
are separated by a partial trace. This theory satisfies the basic
requirements of Sec. \ref{sub:Basic-Requirements} and also always
defines a unique non-zero $\rho_{f}$ %
\footnote{The proof that $\rho_{f}$ is always non-zero follows similarly to
the previous case, by proving that the integrand is always positive
semi-definite and there always exist CV states $|\phi\rangle$ for
which both the integrand and $\mathrm{d}[\phi]$ are non-zero.%
}, thus avoiding both dynamical consistency and information paradoxes.
It is not, however, renormalisation non-linear but polynomial non-linear.

Finally, consider the modification to Eq. (\ref{eq:pureTICTCs}) where
mixed CV states are integrated over, rather than pure CV states. This
similarly satisfies the basic requirements of section \ref{sub:Basic-Requirements},
and also always defines a unique non-zero $\rho_{f}$ %
\footnote{Subject to reasonable assumptions about the measure used.%
}. It is also polynomial non-linear. 

There are, of course, further variations which could be considered.
For example, one could use an alternative generalisation of transition
probability for mixed states, as discussed in Sec. \ref{sub:A-Selection-of}.

The purpose of this discussion is to show that whilst T-CTCs were
concentrated on above, the other theories mentioned in Sec. \ref{sub:A-Selection-of}
also have reasonable properties and may be worthy of further development.

\section{Discussion\label{sec:Discussion}}

Non-linear extensions of quantum theory are subtle since the long-standing
plurality of co-existing interpretations is broken. When considering
time travel this manifests itself in two ways. The first is in the
development and motivation of various possible theories: Ontological
bias will affect decisions made. The second is in using those theories:
Mixed states with ontological differences but the same density operator
may behave differently, as discussed in Sec. \ref{sub:Mixed-states-and}.
Neither of these issues arise when considering time travel classically,
since ontology is generally clear and non-linear evolutions are commonplace.

This uniquely quantum issue has both positive and negative effects
on the theories. The way in which quantum theory works allows theories
of time travel that do not suffer from the paradoxes that are present
classically, but which generally break some of the central structure
of quantum theory. Distinguishability of non-orthogonal states, state
cloning/deleting, and the spectre of superluminal signalling all present
themselves. It also appears that computational power is greatly increased
even beyond that of quantum computers.

Having set out the current state of research into the quantum circuit
approach to time travel, the shortcomings of D-CTCs and P-CTCs are
summarised in Sec. \ref{sub:Desirable-Features}. Most troubling is
that both D-CTCs and P-CTCs suffer from paradoxes (of the information
and dynamical consistency types respectively), and whilst both could
perhaps be eliminated by arbitrarily small noise the theories themselves
remain paradoxical. The two classes of new theories presented in Sec.
\ref{sub:A-Selection-of} were designed to avoid these paradoxes,
and hopefully also satisfy many of the other desiderata of Sec. \ref{sub:Desirable-Features}.

Of the new theories, that of T-CTCs has been selected due to its physical
motivation. To illustrate the strength of the physical story told
in Sec. \ref{sub:T-CTCs}, consider applying the same reasoning in
a classical context.

A CR observer watching a classical time travel circuit sees a CV system
in an unknown ontic state $\tilde{\tau}_{i}$ emerge from the future,
interact with a CR system in a known state, and then disappear back
to the past in the ontic state $\tilde{\tau}_{f}$. For each $\tilde{\tau}_{i}$,
the observer knows that when it heads back in time it must be found
to be in the same state. Whilst in quantum mechanics the probability
of finding a system in a given state is given by the transition probability,
the corresponding probability classically is either unity or zero:
Either $\tilde{\tau}_{i}=\tilde{\tau}_{f}$ or $\tilde{\tau}_{i}\neq\tilde{\tau}_{f}$.
So when the CR observer takes a probability distribution over all
possible CV states $\tilde{\tau}_{i}$, the only states to which non-zero
probabilities are assigned are precisely those states for which $\tilde{\tau}_{i}=\tilde{\tau}_{f}$
after the interaction. What is missing from this account is a way
of specifying the exact probabilities assigned to each CV state. One
might choose to use the principle of indifference and weight each
possibility equally, but this in not necessary.

A very similar argument could be used to argue in favour of D-CTCs,
for example, by demanding exact equality of reduced density operators
rather than consistency via the transition probability. In this case,
using the principle of indifference would lead to a result that is
equivalent to the uniform weighted D-CTCs mentioned in Sec. \ref{sub:A-Selection-of}.
However, by accepting the interpretation of transition probabilities,
and supposing that only pure states are primitive, T-CTCs do have
a clear physical motivation.

Of course, this not a cast-iron argument for T-CTCs, and there is
a certain amount of vagueness in the description given in Sec. \ref{sub:T-CTCs}.
Similar interpretational vagueness is found with both D-CTCs and P-CTCs,
and should be expected when attempting to extend quantum mechanics,
which lacks consensus on interpretation, to a non-linear regime that
is so alien to it.

These arguments for the classical model and uniform weighted D-CTCs
differ from the argument for T-CTCs in an important respect. The probabilities
assigned to the different possible histories with T-CTCs are physically
determined: They are proportional to the transition probabilities.
On the other hand, in the above discussion of classical model and
D-CTCs using the same narrative there is no physical assignment of
probabilities. The use of principle of indifference is an epistemic
move, not a physical one.

Consider the desirable features listed in Sec. \ref{sub:Desirable-Features}
in the light of the theory of T-CTCs. The above argument aims to satisfy
feature 1. As with D- and P-CTCs, feature 2 is satisfied so long as
an appropriate ontology of measurement is chosen (Sec. \ref{sub:Signalling}).
Features 3, 4, and 5 are satisfied without condition, as discussed
in Sec. \ref{sub:T-CTCs}. Feature 6 is partially satisfied, in that
both $\tau$ and $\rho_{f}$ are considered ontologically pure, but
since a proper mixture is taken over so many possibilities, the mathematical
form of either is very rarely pure. In Sec. \ref{sub:Mixed-States-and}
it is shown that feature 7 is satisfied, although related questions
on distinguishing non-orthogonal states with greater-than-quantum
fidelity and broadcasting of mixed states are left open.

Does this mean that T-CTCs are a better model for quantum mechanics
with time travel or, more specifically, for quantum mechanics in the
presence of CTCs? Not necessarily. The physical motivation for T-CTCs
is far from a ``first principles'' argument and there is still the
question as to how, and when, the state projection occurs. However,
both D-CTCs and P-CTCs have incomplete physical motivations and both
leave questions as to how, and when, exactly a proposed physical change
occurs---all three theories share the unobservable dynamical ambiguity.

The theory of D-CTCs is weakened by the failure of the equivalent
circuit model, as currently presented in the literature, to hold up
to scrutiny since that model claimed to provide motivation for the
D-CTC consistency condition and the maximum entropy rule (although
it may still be possible to correct this model). However, the discussion
of Sec. \ref{sub:Non-uniqueness-Ambiguity-in} shows that there is
evidence that the maximum entropy rule may nonetheless arise as a
result of noise. It is an interesting open problem to see if this
can be shown to be true generally, but it would not rid the theory
of the fact that the uniqueness ambiguity is essentially present.

What has been comprehensively shown is that there is a whole landscape
of other theories out there. The quantum circuit approach to quantum
mechanics with time travel may be very attractive in that it abstracts
away from knotty problems with spacetime geometry or any other exact
mechanism for time travel, but it is perhaps too general for the problem
at hand. In order to identify a more robustly physical solution to
quantum mechanics with time travel it may be necessary to use a different
approach, such as path integral or field theoretic ideas. Alternatively,
by very carefully committing to a specific ontology for quantum mechanics,
it may be possible to identify the corresponding theory of time travel.
When non-linearity is present vagueness on this point is problematic.
\begin{acknowledgments}
I would like to thank Jonathan Barrett for helpful discussions and
advice, as well as Jacques L. Pienaar and Mark M. Wilde for useful
comments on an earlier version of this paper. I would also like to
thank the anonymous referees for their comments and criticisms. This
work is supported by the Engineering and Physical Science Research
Council and also by an FQXi Large Grant ``Time and the Structure
of Quantum Theory''. Parts of sections \ref{sec:Introduction}-\ref{sec:D-CTCs-and-P-CTCs}
(notably the review of previous work and the discussion of the equivalent
circuit model) derive from an essay submitted towards Part III of
the Mathematical Tripos at the University of Cambridge for which I
would like to thank Richard Jozsa for his guidance and acknowledge
support from Corpus Christi College, Cambridge. 
\end{acknowledgments}

\bibliographystyle{apsrev4-1}
\bibliography{references}

\end{document}